# Topology optimization of heat sinks for instantaneous chip cooling using a transient pseudo-3D thermofluid model

Tao Zeng[a][*], Hu Wang[a][Ψ], Mengzhu Yang[a], Joe Alexandersen[b]

[a] *State Key Laboratory of Advanced Design and Manufacturing for Vehicle Body, Hunan University, Changsha, 410082, PR China*

[b] *Institute of Technology and Innovation, University of Southern Denmark, Odense, Denmark*

**Abstract:** With the increasing power density of electronics components, the heat dissipation capacity of heat sinks gradually becomes a bottleneck. Many structural optimization methods, including topology optimization, have been widely used for heat sinks. Due to its high design freedom, topology optimization is suggested for the design of heat sinks using a transient pseudo-3D thermofluid model to acquire better instantaneous thermal performance. The pseudo-3D model is designed to reduce the computational cost and maintain an acceptable accuracy. The model relies on an artificial heat convection coefficient to couple two layers and establish the approximate relationship with the corresponding 3D model. In the model, a constant pressure drop and heat generation rate are treated. The material distribution is optimized to reduce the average temperature of the base plate at the prescribed terminal time. Furthermore, to reduce the intermediate density regions during the density-based topology optimization procedure, a detailed analysis of interpolation functions is made and the penalty factors are chosen on this basis. Finally, considering the engineering application of the model, a practical model with more powerful cooling medium and higher inlet pressure is built.

[*] First author: T. Zeng. E-mail: zengtao123@hnu.edu.cn
[Ψ] Corresponding author: H. Wang, E-mail: wanghu@hnu.edu.cn

The optimized design shows a better instantaneous thermal performance and provides 66.7% of the pumping power reduction compared with reference design.

**Nomenclature**

| | |
|---|---|
| $A$ | area |
| $c$ | specific heat |
| $C$ | ratio of properties of fluid phase and solid phase |
| $Da$ | Darcy number |
| $d_{min}$ | minimum mesh size |
| $E$ | pumping energy consumption |
| $ES$ | pumping energy consumption rate |
| $f$ | objective function |
| $\mathbf{f}$ | source vector term in Navier-Stokes equations (2) |
| $f_b$ | heat source of base plate |
| $f_c$ | source term of energy conservation equation (4) |
| $f_v$ | volume friction |
| $h$ | heat exchange coefficient |
| $I$ | interpolation function |
| $I_s$ | interpolation function of product of density and specific heat interpolation functions |
| $k$ | thermal conductivity |
| $L_c$ | characteristic length |
| $\mathbf{n}$ | unit normal vector |
| $n_{ev}$ | maximum evaluation number |
| $p$ | pressure |
| $p_{drop}$ | pressure drop |
| $P_{pump}$ | pumping power |
| $q$ | penalty factors |
| $q_{bp,d}$ | heat dissipating power directly from base plate to air |
| $q_{fin}$ | heat dissipating power from heat sink fins to air |
| $q_{in}$ | inner heat transferred between the channel layer and base plate layer |
| $q_{out}$ | heat flux through the outlet |
| $q_s$ | penalty factors of product of density and specific heat interpolation functions |
| $Q$ | rate of heat production |
| $r_{fil}$ | filter radius |
| $r_{min}$ | minimum radius |
| $r_f$ | volumetric flow rate through inlet |
| $s$ | product of density and specific heat |
| $t$ | time |
| $t_{bp}$ | thickness of heat sink base plate |
| $t_{ch}$ | thickness of heat sink channel |
| $T$ | temperature vector |
| $T_b$ | temperature of base plate |
| **u** | velocity vector |
| $v$ | velocity magnitude |
| $V$ | volume |
| **x** | system coordinate vector |

*Greek symbols*

| | |
|---|---|
| $\alpha$ | inverse permeability |
| $\beta$ | projection steepness parameter |
| $\bar{\alpha}$ | maximum inverse permeability |
| $\gamma$ | residual of the finite element formulation |
| $\varepsilon$ | tolerance of optimization process |
| $\rho$ | material density |
| $\theta$ | design variable |
| $\theta_0$ | initial value of design variable |
| $\mu$ | dynamic fluid viscosity |
| $\psi$ | field variables vector |
| $\Omega$ | domain |
| $\Gamma$ | domain boundary |
| $\eta$ | projection threshold parameter |
| $\bar{\bar{\theta}}$ | projected design field |

*Subscripts*

| | |
|---|---|
| $f$ | fluid phase |
| $c$ | cooling procedure |
| $ht$ | heating procedure |
| $s$ | solid phase |
| $in$ | inlet |
| $out$ | outlet |
| $bp$ | base plate |
| $wet$ | regions where solid material is in contact with air |
| $air$ | air |
| $fin$ | heat sink fins |
| $T$ | terminal point |
| $\alpha$ | inverse permeability |
| $k$ | thermal conductivity |
| $h$ | heat exchange coefficient |
| $d$ | design (domain) |
| $avg$ | average value of base plate (or chip) |
| $hs$ | heat sink |
| $opt$ | optimized model |
| $reg$ | ordinary model |

## 1. Introduction

Thermal management of electronics is becoming more and more challenging with the advancement of chips in miniaturization and performance. While two-phase cooling can reach extreme heat fluxes, its application is limited by the complexity of fabrication, assembly and operation [1]. For instance, recent gaming central-processing-units (CPUs) such as Intel® Core<sup>TM</sup> i9-9900K are very popular among consumer chips. It has 8 cores (16 threads) and a 3.6GHz base frequency, but more importantly, it can overclock to over 5GHz. Although the gaming performance improves a lot, its Thermal Design Power (TDP) has reached 95W in a 37.5*mm*×37.5*mm*×1.15*mm* chip size, which brings challenges to heat dissipation. According to recent surveys in Intel® official website [2]，the chip turbo frequency condition has a limit temperature (junction temperature). When the temperature exceeds this temperature limit, the chip frequency will be actively reduced along with the chip's performance. To minimize the time of chip frequency reduction, an active cooling device is proposed to achieve the minimum temperature of the chip below the turbo temperature limit as fast as possible.

There exist two common types of active coolers: air-cooled heat sink and liquid-cooled heat sink. Liquid heat sinks are prohibitive for many consumers due to their high price and short service life. However, when using air-cooled heat sinks, it is hard to satisfy the heat dissipation requirement of high-performance chips like the Intel® Core<sup>TM</sup> i9-9900K. One of the efficient methods to improve the thermal performance of heat sink is structural optimization.

For the structural optimization of the chip heat sink, previous researchers proposed an oblique fin design [3] and wavy channel designs [4,5] for heat sinks. Bejan and Errera et al. [6] proposed convective trees of fluid channels. Another design approach inspired by natural structures is fractal-like flow networks: Taylor et al. [7] utilized the fractal-like branching in microchannel heat sink to reduce the pumping power and wall temperature in the system; Chen et al. [8] were inspired by the fractal pattern of mammalian circulatory and respiratory systems, and they also made a comparison of the new design with the traditional parallel net. Recently a novel concept for energy efficiency hotspot targeted liquid cooling of microprocessors was proposed by Sharma et al. [9] and they greatly reduced chip temperature non-uniformities. In addition to the new design of fins and channels and intuitive structures inspired by natural structures, shape optimization was implemented to multi-objective optimization of a heat exchanger with parallel genetic algorithms by Hilbert et al. [10].

Topology optimization is a higher-level structural optimization method originally introduced to design optimal topologies with a homogenization method by Bendsøe et al. [11]. With the development of computer technology and numerical calculation, there are now many branches of topology optimization methods [12]. One of the most widely-used topology optimizations is the Solid Isotropic Material with Penalization (SIMP) approach [13,14]. The SIMP method has been widely deployed on many Computer Aided Engineering (CAE) systems, such as ANSYS, ABAQUS and so on. It is now generally known as the "density-based" approach and because of its versatility and expandability, the method is now applied in many fields, such as heat transfer, fluids, optics and acoustic [12]. However, the elimination of intermediate density elements has always been and is still an important topic in the density-based topology optimization method. Another method of topology optimization, the level set method, has the advantage of having clearly defined phases and geometrically smooth and clear boundaries. It was first proposed by Wang et al. [15] and the level set method has been extended for continuum structure by Allaire et al. [16] and Xia et al. [17,18]. Although the level set can provide a clear definition of the boundary, it is only captured in the physics when combining with a boundary conforming method such as the extended finite element method (X-FEM) [19]. Xie et al [20] proposed Evolutionary Structural Optimization (ESO) method, which avoids intermediate density elements using discrete updates based on intuitive stress limit condition. However, the ESO method has some difficulties with convergence and expandability, but these are, to some degree, reduced using the Bi-directional Evolutionary Structural Optimization (BESO) method [21]. With the development of recent topology optimization technology, it is widely used in mechanical structures, heat transfer and fluid problems amongst others [22].

Thermofluid topology optimization is an important branch of heat sink optimization because it provides a scheme for optimizing the temperature field as well as the fluid flow simultaneously, which affect the heat dissipation performance of heat sinks. However, thermofluid problems require the simulation of both the fluid flow and the heat transfer, coupling the fluid flow to the temperature field through convection. The computational cost of a single simulation may be an obstacle to optimization, requiring hundreds of simulations due to its iterative nature. In order to overcome that, many researchers have used Newton's law of cooling combined with a constant heat transfer coefficient to approximate the heat transfer to a fluid. Yin et al. [23] proposed a novel topology design scheme for electro-thermally actuated compliant mechanism. Another approximate thermofluid model method is to use a surrogate model, that has been used by Iga et al. [24] and Joo et al. [25]. Bruns [26] investigated topology optimization of convection-dominated, steady-state heat transfer, proposing interpolating the convection boundary using density variations from element to element. This approach has been applied by [27] and [28] and was recently formalized in a continuous formulation using density-gradients by [29] and [30]. However, recent advances in computational power now allows for optimization of more complex problems. Therefore, many researchers are beginning to pursue an accurate solution to thermofluid topology optimization using full conjugate heat transfer models. Moreover, forced convection was initially investigated by [31] and [32] and has subsequently been extended by many authors, as is summarized in the review paper by [33]. Recently, turbulent fluid flow [34] and forced heat transfer [34] has been presented using a density-based topology optimization approach. Natural convection is less studied, with Alexandersen et al. [35] treating it for the first time using a 2D model. Subsequently, Alexandersen et al. [36] proposed a large scale fully parallel computational framework as a way to topology-optimize high-fidelity 3D heat sinks cooled by natural convection as well as passive cooling of light-emitting diode lamps [37]. In order to decrease the computational cost, Joo et al. [38] proposed a simplified model using Newton's law of cooling and correlations for the same problem. Furthermore, a simplified potential flow model has recently been proposed as a way to reduce computational cost at an acceptable accuracy [39,40].

Although topology optimization of 2D thermofluid models to some extent can predict the optimized shape for forced convection problems, it has inevitable error compared with full 3D thermofluid topology optimization. However, as pointed out above, this comes at a high computational cost and time. Therefore, many researchers are exploring simplified 2D approximations of the full 3D problems, giving a cheaper computational cost with acceptable. A pseudo-3D thermofluid model was first proposed by Haertel et al. [41] coupling a solid thermal base layer to a fluid-solid cross-sectional flow layer. In extension of this work, Zeng et al. [42] used a similar model to topology-optimize a forced air heat sink with superior heat sink performance investigated through experimental and numerical investigation. Recently, Yan et al. [43] assumed a fourth-degree polynomial temperature profile of the heat sink thermal-fluid layer and a linear temperature profile in the substrate to do topology optimization at close to 2D computational cost with increased accuracy. All of the mentioned works about topology optimization of thermofluid model are steady-state problems. Therefore, in view of the operating conditions of chip heat sinks, topology optimization of transient thermofluid is necessary, especially for treating the instantaneous behavior that requires to cool the chip as fast as possible. Therefore, a transient pseudo-3D thermofluid model is presented in this work.

This paper is organized as follows: the transient pseudo-3D forced convection heat sink model is developed in Section 2; the verification of pseudo-3D model is covered in Section 3, by comparing the pseudo-3D model with real 3D model for steady-state conditions; the overall topology optimization formulation is illustrated in Section 4 with implementation details; the superior performance of topology-optimized designs are verified through the comparison with a reference chip heat sink in Section 5, including comparison of transient optimized results, steady-state optimized results and a more practical model; finally, discussion and conclusions are provided in Section 6.

## 2. Transient pseudo-3D forced convection heat sink model

A schematic illustration of a full 3D chip heat sink model is shown in Fig. 1. The channel of the model is teemed with cooling medium and three heat sink slices are placed in the middle of the channel. At the bottom of the heat sink, there is a chip, which generates heat. The heat of the chip is transferred mainly from the chip to the slices and then is taken away by the flowing cooling that flows in from inlet and flows out to the outlet. The simulation of full 3D model in this study is very costly not to mention the optimization procedure. Thus, the pseudo-3D model is utilized in this study because it can be obtained by extruding from a 2D model.

As illustrated in Fig. 2, under the full 3D heat sink model is a "channel layer" defined as a cross-section through the full model. To demonstrate its pseudo 3D model more clearly, a pseudo 3D model structure sketch is displayed separately in Fig. 3.

As is shown in Fig. 3, there is a plane coupled to another layer containing the cross section of the base plate. The upper plane "channel layer" couples the temperature and flow fields, where cooling air flows in from the inlet boundary of the channel, flows through and around the heat sink fins and flows out through the outlet boundary. This layer also contains the heat sink fin region, which transfers the heat from the base plate at the bottom to the cooling air. The heat sink fin cross-sectional shape then affects the flow paths of the cooling air, which might greatly influence the heat dissipation performance of the heat sink.

The other region of the layer is a pure fluid region. Two non-slip boundaries are added at two longer edges of the model. Constant pressures are imposed on the inlet and the outlet of the channel. To simulate the chip reaching the turbo limit temperature and the start of the active cooling device, the temperature is set to 90°C, which is considered to be the limit temperature for the i9-9900K chip in this study. The ambient temperature of the model is 25°C, which is also applied as the Dirichlet boundary on the inlet boundary. The lower layer is the "heat source layer", which consists only of the heated base plate. With a uniformly constant heat production. All boundaries of this layer are adiabatic boundaries.

Since the two individual layers are built separately, it is crucial to connect them properly. As disrobed in Ref.[41], one part of the heat production is absorbed by the fins in the channel layer. The other part of heat production causes a rise of temperature of the base plate. The heat absorbed by the fins is much more easily taken away by the cooling medium than directly through the heat source, when the cooling air passes through the channel.

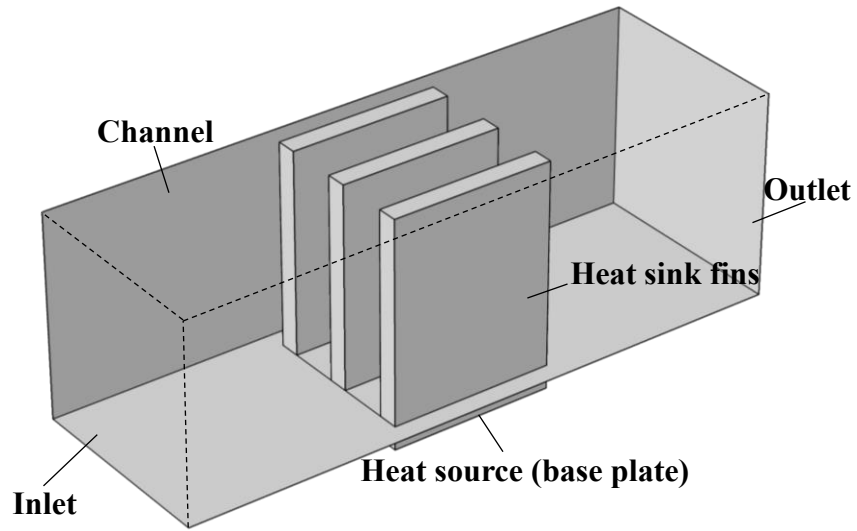

Fig. 1. Full 3D model of a reference straight-fin heat sink.

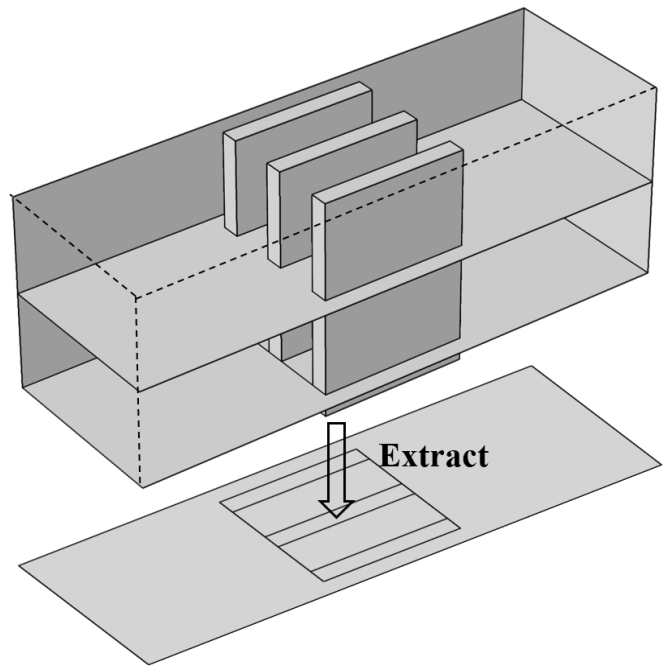

Fig. 2. Illustration of full 3D model and the corresponding solid-fluid "channel layer" of the pseudo-3D model.

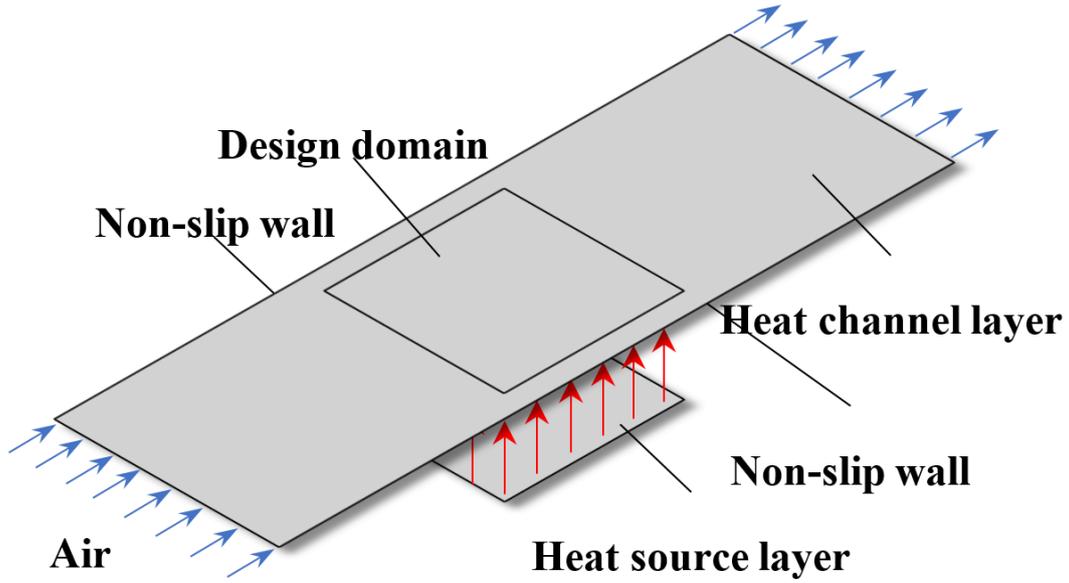

Fig. 3. Transient pseudo 3D model structure sketch

To depict the model more accurately and analyze it subsequently, some assumptions are introduced in this model:

a) The flow field of the channel layer constitutes a laminar and incompressible flow;

b) The material properties of all phases are constant and do not change with temperature;

c) The fluid flow and the heat transfer are analyzed using a time-dependent solver since the instantaneous behavior is of interest.

2.1. Channel layer

**Mathematical model of channel model**

As the channel layer has 2 physical fields, flow field and temperature field, these two fields interact with each other. To establish an effective mathematical model for simulation, the Navier-Stokes equations of the channel layer are formulated as follow:

$$\frac{\partial \rho_f}{\partial t} + \nabla \cdot (\rho_f \mathbf{u}) = 0 \tag{1}$$

$$\rho_f \left( \frac{\partial \mathbf{u}}{\partial t} + \mathbf{u} \cdot \nabla \mathbf{u} \right) = -\nabla p + \mu \nabla^2 \mathbf{u} + \mathbf{f} \tag{2}$$

where $\mathbf{f}$ represents the source term of the fluid flow; $\rho_f$ in Eqs. (1) and (2) denotes the fluid density; $\mathbf{u}$ is the velocity field of the model; $p$ is the fluid pressure; and $\mu$ is the fluid viscosity. Due to the incompressible assumption and, a constant density of the fluid, the mass conservation Eq. (1) changes as follows:

$$\nabla \cdot \mathbf{u} = 0 \tag{3}$$

When it comes to the heat transfer and the fluid flow, the energy conservation equation usually contains several terms representing different physical meanings, such as: heat diffusion term $k\nabla^2 T$; convection term $\rho_f c_f \mathbf{u} \cdot \nabla T$; temporal term $\rho_f c_f \partial T / \partial t$; and a source term $f_c$. Thus, synthesizing all differential terms above, the energy conservation equation is formulated as:

$$\rho_f c_f \frac{\partial T}{\partial t} + \rho_f c_f \mathbf{u} \cdot \nabla T - k\nabla^2 T = f_c \tag{4}$$

where $c_f$ is the specific heat of the fluid and $T$ denotes the temperature field of the model.

**Boundary conditions of channel layer**

As for the temperature boundary conditions of the channel layer, the fluid inlet temperature is set to the value of $T_{in} = 25°C$, which can be expressed as:

$$T = T_{in} \quad \text{on} \quad \Gamma_{in} \tag{5}$$

where $\Gamma_{in}$ represents the inlet boundary of the model. The temperature field boundary conditions for the outlet and walls in the channel layer are given by:

$$\mathbf{n} \cdot \nabla T = 0 \quad \text{on} \quad \Gamma_{out} \bigcup \Gamma_{wall} \tag{6}$$

The temperature of regions except the inlet boundary is set to a uniform value at the initial time, which means that the temperature of the chip starts at the turbo frequency temperature limit $T_0 = 90°C$, and it is defined as:

$$T(\mathbf{x},t)\big|_{t=0} = T_0 \quad \mathbf{x} \in \Omega \bigcup \mathbf{x} \notin \Gamma_{in} \tag{7}$$

The velocity and the pressure boundary conditions of the channel layer are as follows:

$$\begin{cases} p = p_{in} & \text{on } \Gamma_{in} \\ p = 0 & \text{on } \Gamma_{out} \end{cases} \quad (8)$$

The magnitude of heat source term $f_c$ of Eq. (4) consists of two parts. One part comes from the base plate heat production that is transferred to the channel layer through fins of the heat sink. The other part is from the cooling airflow that passes through the heat sink fins region in the channel layer. Therefore, the heat source term $f_c$ is defined as:

$$f_c = \frac{h(T_b - T)}{t_{ch}} \quad (9)$$

$$h = \begin{cases} h_s, & \text{Solid phase} \\ h_f, & \text{Fluid phase} \end{cases} \quad (10)$$

where $h$ is the convection coefficient of the channel layer, $T_b$ is the base plate temperature field, and $t_{ch}$ is the thickness of the channel layer.

2.2. Base plate layer

**Mathematical model of base plate**

The base plate is composed of the solid phase material, whose material property is constant and does not change with temperature. A constant heat production rate is imposed on the base plate. However, most of the heat is transferred to the channel layer. Thus, the differential equation denotes as follow:

$$\rho_s c_s \frac{\partial T_{bp}}{\partial t} - \nabla \cdot (k_s \nabla T_{bp}) = f_b \quad (11)$$

where $\rho_s$, $c_s$ and $k_s$ are the density, specific heat and thermal conductivity of the solid phase material, respectively. $f_b$ represents the heat source of the base plate and it can be expressed as the difference between heat production and heat dissipation:

$$f_b = \frac{Q}{V_{bp}} - \frac{q_{in}}{t_{bp}} \quad (12)$$

where $Q$ represents the heat production power, $V_{bp}$ is the volume of material of base plate, $t_{bp}$ is the thickness of the base plate and $q_{in}$ denotes the inner heat transferred between the channel layer and base plate layer, defined as:

$$q_{in} = h(T_{bp} - T) \quad (13)$$

**Boundary conditions of base plate**

The boundaries of the base plate are adiabatic, which is expressed as:

$$\mathbf{n} \cdot \nabla T_b = 0 \text{ on } \partial\Omega_b \quad (14)$$

2.3. Determination of artificial heat convection coefficient $h$

The heat exchange in the full 3D model conforms regularity of heat transfer on the interface between solid and fluid, and the special heat transfer inside of the solid and fluid. The pseudo-3D model, however, only provides the heat transfer along with the layers without the heat conduction perpendicular to the layer. Therefore, the artificial heat convection coefficient is introduced to measure the heat flux intensity between the two layers whose direction is vertical to the layer. Obviously, the value of the heat convection coefficient in the pseudo-3D model has a great influence on the accuracy of the model, which means that it will determine whether the pseudo-3D model could replace the full 3D model accurately during the simulation and optimization procedure. Haertel et al. choose the value of $h$ intuitively in [41]. However, Zeng et al. provided detailed derivation processes in [42], which is used as a reference in this work.

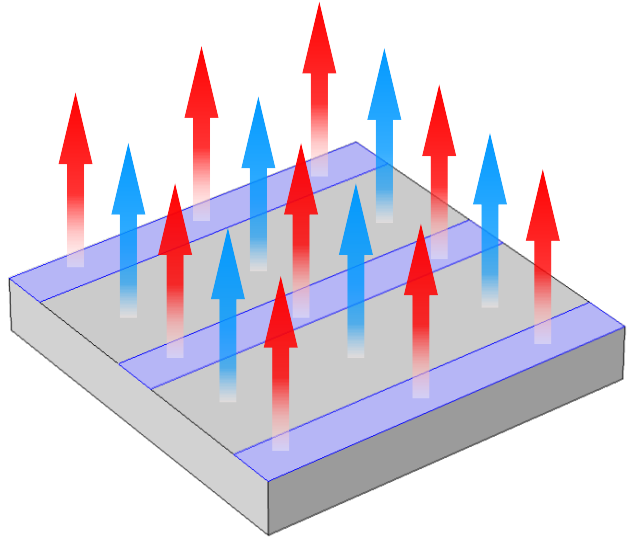

Fig. 4. Two main conduction paths of heat transfer from the heat source

The heat dissipation process of the 3D model shown in Fig. 1 contains two parts as shown in Fig. 4:

(1) Heat transferred to the air through the fins of the heat sink, shown as red arrows;

(2) Heat transferred directly to the air through the base plate exposed to the air, shown as blue arrows.

The two different areas have their heat dissipation capacity, which represents that they have different values of $h$. Therefore, it takes two limit values, representing the ability to dissipate heat in the solid, denoted as $h_s$, and in the fluid, denoted as $h_f$.

For the fluid region, $h_f$ reflects the heat convection ability of the base plate material with air. The definition is fully based on the numerical solution of the full 3D model in Fig. 1, and is given as:

$$h_f = \frac{q_{out}}{A_{wet}(T_{wet} - T_{air})} \tag{15}$$

$$q_{out} = \int_{\Gamma_{out}} \rho_f c_f u(T - T_{in}) d\Gamma \tag{16}$$

where $q_{out}$ is the heat power brought out from the channel by cooling air at the outlet. $A_{wet}$ and $T_{wet}$ represent the area and average temperature of the surface of solid material contact with air, respectively. $T_{air}$ in Eq. (15) is the average temperature of the air close to the solid surface.

The second part of the heat power is defined as follow:

$$q_{bp,d} = h_f A_{bp,wet}(T_{bp,wet} - T_{air,bp}) \quad (17)$$

where $A_{bp,wet}$ and $T_{bp,wet}$ are the area and the temperature of the surface of base plate exposed to cooling air, respectively. $T_{air,bp}$ represents the average temperature of the air close to the base plate. Therefore, the first part of heat dissipation power can be obtained by the difference between $Q$ with $q_{bp,d}$:

$$q_{fin} = Q - q_{bp,d} = Q - h_f A_{bp,wet}(T_{bp,wet} - T_{air,bp}) \quad (18)$$

The heat conduction capacity from base plate to the heat sink fins is defined as:

$$h_s = \frac{q_{fin}}{A_{fin,bp}(T_{bp} - T_{fin})} \quad (19)$$

where $A_{fin,bp}$ is the area of region of base plate, with which the fins are contact, $T_{bp}$ and $T_{fin}$ are the average temperature of the base plate and fins, respectively. Since all the values above are transient, the determination of $h_s$ and $h_f$ is calculated by using the average value during a certain time period.

## 3. Validation of transient pseudo 3D model

Although the pseudo-3D model is computationally cheaper than the full 3D model, it is an approximate model and its accuracy needs to be verified. The measure that is used to compare the two models is an average temperature criterion of the base plate:

$$f(t) = \begin{cases} \dfrac{1}{A_{bp}} \int_{\Omega_{bp}} T_{bp} - T_{in} d\Omega, & \text{for pseudo-3D} \\ \dfrac{1}{V_{bp}} \int_{\Omega_{bp}} T_{bp} - T_{in} d\Omega, & \text{for 3D} \end{cases} \quad (20)$$

The comparative benchmark is a 3D finite element (FE) model with three uniform straight fins inside of the heat sink channel. The temperature distribution of the base plate layer in the transient pseudo-3D model and the heat generation domain at the bottom of the channel will be used for comparison.

In the implementation of the transient pseudo-3D model, the differential equations of the overall model can be expressed as:

$$\begin{cases} \nabla \cdot \mathbf{u} = 0 \\ \rho_f \left( \dfrac{\partial \mathbf{u}}{\partial t} + \mathbf{u} \cdot \nabla \mathbf{u} \right) = -\nabla p + \mu \nabla^2 \mathbf{u} + \mathbf{f} \\ \rho_f c_f \dfrac{\partial T}{\partial t} + \rho_f c_f \mathbf{u} \cdot \nabla T - k \nabla^2 T = \dfrac{h(T_{bp} - T)}{t_{ch}} \\ \rho_s c_s \dfrac{\partial T_{bp}}{\partial t} - \nabla \cdot (k_s \nabla T_{bp}) = \dfrac{Q}{V_{bp}} - \dfrac{h(T_{bp} - T)}{t_{bp}} \end{cases} \quad (21)$$

The finite element analysis (FEA) procedure is implemented in COMSOL Multiphysics 5.4. The physical field variables in this model are composed of the velocity field $\mathbf{u}$, pressure field $p$, temperature field of the channel layer $T$ and temperature field of the base plate layer $T_{bp}$.

The dimensions of the model are shown in Fig. 5. The parameters and material properties of the model are displayed in **Table 1** and **Table 2**. The values of $h_s = 9.8 \times 10^4$ W/(m²·K) and $h_f$ = 90 W/(m²·K) are obtained by the calculation of the full 3D model according to Eqs. (15) - (19). Aluminum is chosen as the material of the solid phase of the model, with the material properties shown in **Table 2**. The inlet pressure of both models is set to 2Pa. Moreover, the mesh size ranges of pseudo 3D model and full 3D model are 0.3-0.45*mm* and 0.54-3*mm*, and their computation time are 32*s* and 1,585*s*, on the same computer with Intel® Xeon™ E3-1230 V2 CPU and 16GB RAM. Obviously, just a single simulation procedure of pseudo-3D model can cut down much computational expense.

**Table 1**

Parameters of boundary conditions and model property of the pseudo 3D model and 3D model.

| Parameters | Pseudo 3D model | 3D model |
|---|---|---|
| $T_{in}$ [℃] | 25 | 25 |
| $t_{ch}$ [mm] | 10 | 10 |
| $t_{bp}$ [mm] | 1.15 | 1.15 |
| $Q$ [W] | 4.3236 | 4.3236 |
| $p_{in}$ [Pa] | 2 | 2 |
| $p_{out}$ [Pa] | 0 | 0 |
| $h_s$ [W/(m²·K)] | 9.8×10⁴ | — |
| $h_f$ [W/m²·K] | 400 | — |

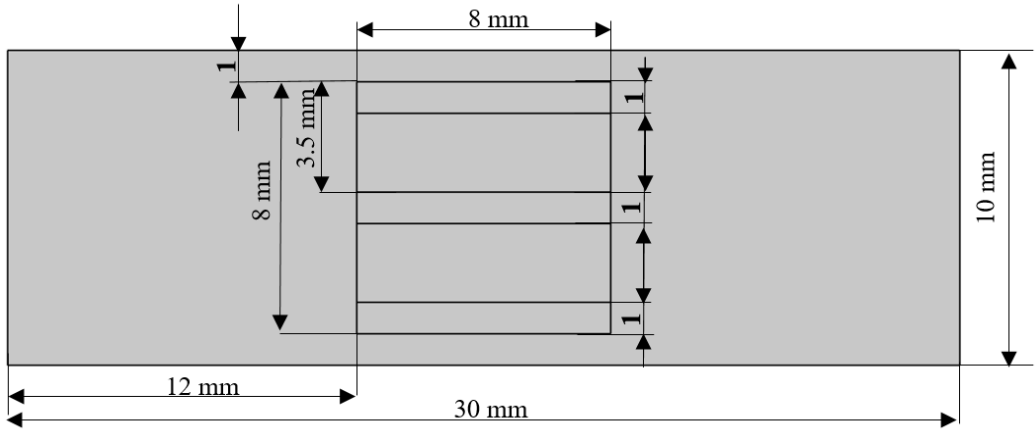

Fig. 5. Geometric sizes of transient pseudo 3D model

**Table 2**

Thermo-physical properties of pseudo 3D model.

| Thermo-physical properties | Values |
|---|---|
| $k$ [W/(m·K)] | 0.024 |
| $k_s$ [W/(m·K)] | 237 |
| $\rho_f$ [kg/m³] | 1.204 |
| $\rho_s$ [kg/m³] | 2,700 |
| $c_f$ [J/(kg·K)] | 1006 |
| $c_s$ [J/(kg·K)] | 900 |
| $\mu$ [Pa·s] | 1.94×10⁻⁵ |

With the FEA of transient pseudo-3D and full 3D model implemented in the software, the measure $f(t)$ is compared over time. To guarantee the accuracy of the transient process of heat dissipation, a relatively long period time [0, 100]$s$ is analyzed by the two models assuring that the field variables transit from transient to steady-state. As shown in Fig. 6, the measure of the two converge to nearly the same value, with only minor differences during the period time. A single indicator may not fully explain the equivalence of the two models, so the distribution of the temperature field of the base plate of the two models can help to explain its effectiveness and accuracy. To further describe their similarity in Fig. 7 quantitatively, a criterion $g$ is introduced:

$$g = \frac{1}{A_{bp}} \int_{A_{bp}} \frac{|T_{3D} - T_{p3D}|}{T_{3D}} dA \tag{22}$$

where $T_{3D}$ and $T_{p3D}$ represent the temperature field of the base plate of the 3D and pseudo-3D models, respectively. The temperature field plane of the full 3D is the intermediate section of the chip locating at the height of 0.575*mm*.

A value of $g = 2.462 \times 10^{-3} < 1\%$ is obtained, which represents highly similar temperature fields between the two models. That means the transient pseudo-3D model could be used as the approximate replacement model, which facilitates the simulation and later topology optimization.

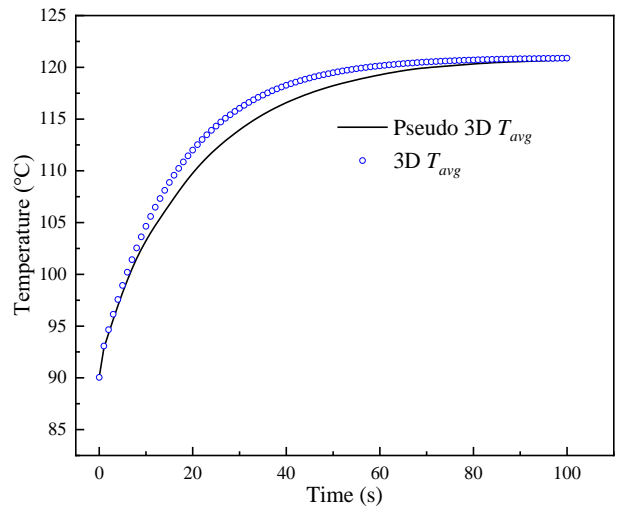

Fig. 6. Average temperatures of base plate of pseudo 3D and 3D models.

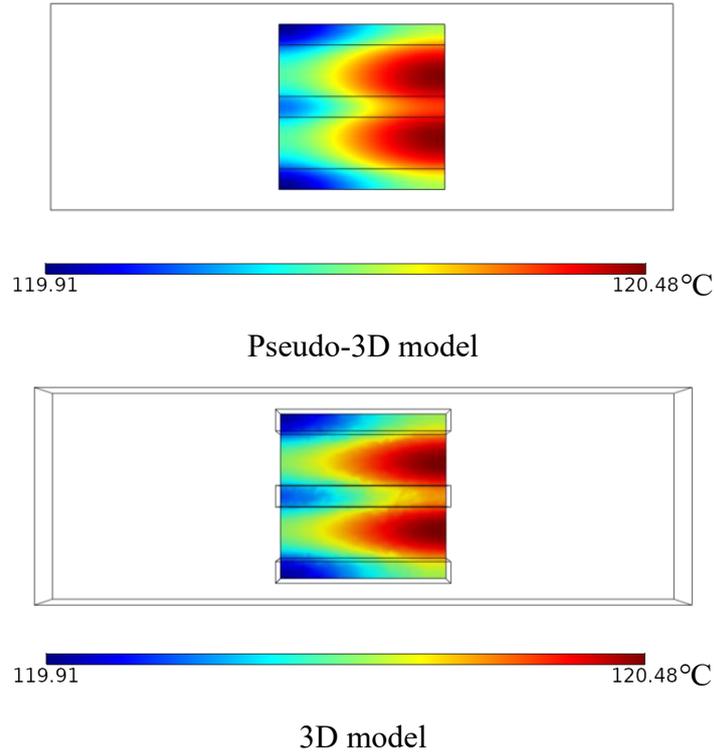

Pseudo-3D model

3D model

Fig. 7. Temperature distributions of the base plate of pseudo-3D and 3D models at 100*s*.

## 4. Topology optimization

With the successful verification of the transient pseudo-3D model, topology optimization is now introduced to obtain superior heat dissipation structures. The design variable $\theta$ is introduced in the design domain of the channel layer and varies in the range of $(0,1]$, and affects the material property of the model.

4.1. Modified governing equations

Since the values of design variables represent the material phase of the design domain, the separation of the governing equations for the two phases in the design domain is important. In this study, a Brinkman friction term is added to the Navier-Stokes equations. The Brinkman friction term is used in fluid flow topology optimization to penalize flow through solid areas within the design domain and corresponds to the force exerted on a fluid flowing through an ideal porous medium [44]. Thus, the Brinkman friction term is defined as:

$$\mathbf{f} = \alpha(\theta)\,\mathbf{u} \tag{23}$$

where $\alpha(\theta)$ represents the Brinkman friction coefficient and it can be expanded as:

$$\alpha(\theta) = \bar{\alpha} \cdot I_\alpha(\theta) = \frac{\mu}{DaL_c^2} \cdot I_\alpha(\theta) \tag{24}$$

where $I_\alpha(\theta)$ represents the interpolation function of the Brinkman friction coefficient as a function of the design variable. $Da$ is the Darcy number of the fluid and $L_c$ denotes the characteristic length of the model. In this work, $Da$ is set to $1\times10^{-5}$ and $L_c$ is equal to 10mm, which is also the height of the 3D channel.

In the energy conservation equation (4) of the channel layer, there are several material properties including material density $\rho$, specific heat $c$, thermal conductivity $k$ and convection coefficient $h$. As seen in Eq. (4) and Eq. (11), $\rho$ and $c$ always appear in the form of a product of both. To minimize the complexity of material interpolation for optimization, this product is considered as an independent material parameter $s=\rho c$ in the optimization procedure. Thus, the energy conservation equation can be modified as:

$$s(\theta)\frac{\partial T}{\partial t} + s(\theta)\mathbf{u}\cdot\nabla T - k(\theta)\nabla^2 T = \frac{h(\theta)(T_{bp}-T)}{t_{ch}} \tag{25}$$

where $s(\theta)$, $k(\theta)$ and $h(\theta)$ represent the corresponding material properties with respect to design variable $\theta$:

$$\begin{cases} s(\theta) = s_f \cdot I_s(\theta) \\ k(\theta) = k_f \cdot I_k(\theta) \\ h(\theta) = h_f \cdot I_h(\theta) \end{cases} \tag{26}$$

As for the heat transfer equation for the base plate layer, Eq. (10), it is modified as:

$$\rho_s c_s \frac{\partial T_b}{\partial t} - \nabla\cdot(k_s \nabla T_{bp}) = \frac{Q}{V_{bp}} - \frac{h(\theta)(T_{bp}-T)}{t_{bp}} \tag{27}$$

## 4.2. Interpolation functions

A detailed analysis of the interpolation functions is introduced to study the variation of the chosen objective function with the design variables, in order to obtain a better optimization behavior and better performing optimized design. Interpolation function is an important part of density-based topology optimization because it can affect the convergence to local optimum in the final design. The interpolation functions $I_j(\theta), j = \alpha, k, h, s$ of this work are defined as:

$$I_j(\theta) = \begin{cases} \dfrac{1-\theta}{1+q_\alpha \theta}, & j = \alpha \\ \dfrac{\theta(C_j(1+q_j)-1)+1}{C_j(1+q_j \theta)}, & j = k, h, s \end{cases} \tag{28}$$

where $q_\alpha$ and $q_j$ represent the penalty factors of inverse permeability and other material properties. $C_j$ is the ratio of properties of the fluid and solid phases:

$$C_j = \begin{cases} \dfrac{k_f}{k_s}, & j = k \\ \dfrac{h_f}{h_s}, & j = h \\ \dfrac{s_f}{s_s}, & j = s \end{cases} \tag{29}$$

The shape of each interpolation function with different values of penalty factors is shown in Fig. 8.

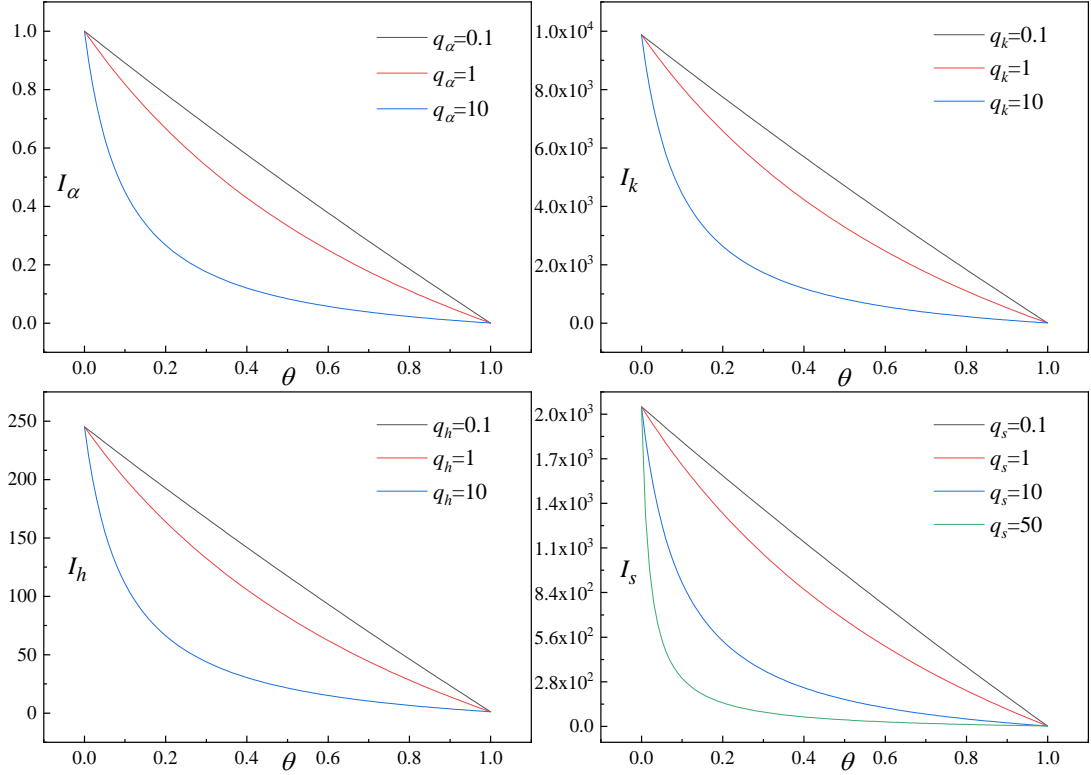

Fig. 8. Interpolation functions of pseudo 3D model

Since the interpolation functions are defined, the proper values of the penalty factors must be selected. In the next section, the importance and the method of the choice of penalty factor values are demonstrated.

4.3. Selection of the penalty factors

The model used to make the choice for penalty factors are shown in Fig. 5. The fin regions are set to a uniform design variable value and the other region is a fluid phase. Before the choice of penalty factors is made, the objective function is defined in order to measure the behavior of varying the penalty factors. The objective function is chosen as:

$$f(\theta) = \frac{1}{A_b} \int_{\Omega_{bp}} T_b - T_{in} d\Omega \Big|_{t=t_T} \qquad (30)$$

where $f(\theta)$ is the relative average temperature of the base plate at the terminal time point. In this study, the terminal time is set to $t_T = 1s$ because optimizing the instantaneous behavior of the heat sink is considered in this work.

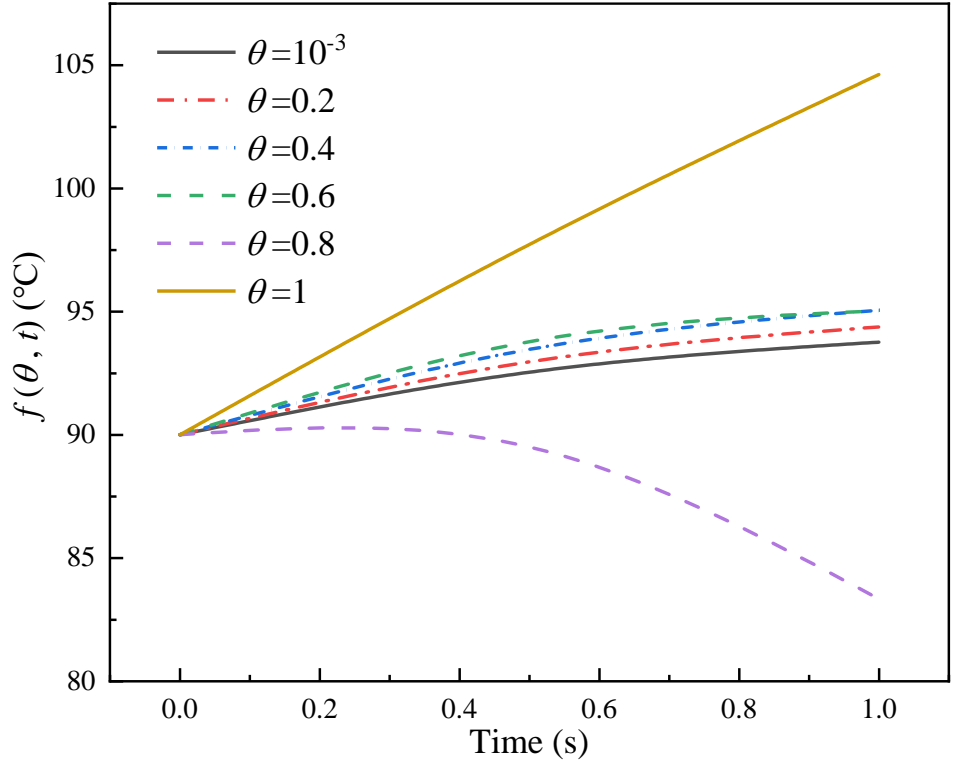

Fig. 9. Objective function varies with respect to time and different design variable values.

In order to demonstrate the value of objective function varying with respect to time and different densities of the model, a figure of $f(\theta,t)$ calculated for the model shown in Fig. 9. The objective function is seen to generally increase with an increasing density value, but at the value of 0.8, the objective function curve is abnormal with a decrease in $f$ over time. This indicates that unphysical behavior is obtained for intermediate design variables values. However, the figure of $f(\theta,t)$ does not show the relation between $f$ and $\theta$ clearly. Since the temperature of the chip at the terminal time point is the focus of this study from now on, a new study to describe the relationship between the objective function $f(\theta)$ shown in Eq. (30) and design variable $\theta$ is implemented.

Similarly to the studies published by [45] and [46], the objective function is shown in Fig. 10 for many sets of penalty factors for varying design variable $\theta$. An intuitive rule can be obtained from the 27 subgraphs with 108 curves in Fig. 10: each penalty factor value has its own special effectiveness on the position of the minimum point of the curve.

In order to reduce the number of intermediate design variables and to converge nicely to an optimized result, the objective function $f(\theta)$ should be a monotonously increasing curve from $\theta=0$ (being fully solid fins) to $\theta=1$, (being no fins or fully fluid). This ensures that the minima with respect to the single variable is at the case of fully solid fins. Furthermore, it ensures a smooth transition from one value to another, without any local minima with respect to the single variable.

The observed trends are as follows:

a) With an increase of the value of $q_\alpha$, the position of the minimum point moves left and downward;

b) With an increase of the value of $q_k$, the position of the minimum point moves left and downward;

c) With an increase of the value of $q_h$, the position of the minimum point moves left and up;

d) With an increase of the value of $q_s$, the position of the minimum point moves right and up.

The above rules may be helpful in finding the values of superior penalty factors, based on their specific sensitivities to the position of the minimum point. They can be used to determine the trends with respect to the penalty factors, but cannot necessarily be used to obtain the exact values of the penalty factors. Therefore, the estimated value of the penalty factors still needs to be verified.

Based on the study and the observed trends, the values of the penalty factors chosen in this study are $q_\alpha=0.1$, $q_k=0.1$, $q_h=50$, $q_s=100$, respectively. To find out whether the chosen parameters of the curve is monotonically increasing, the graph demonstrating the relationship of the objective function with respect to the design variables is individually displayed in Fig. 11. The curve in Fig. 11 shows a monotonically increasing trend, and therefore they represent a good choice of interpolation penalty factors. It should be noted that this study only presents guidance for the choice of penalty factors based on a single global design variable case. The extension to hundreds or thousands of local design variables cannot be guaranteed to follow the exact same trends. However, this provides a qualified choice of the penalty factors rather than choosing them at random or from intuition.

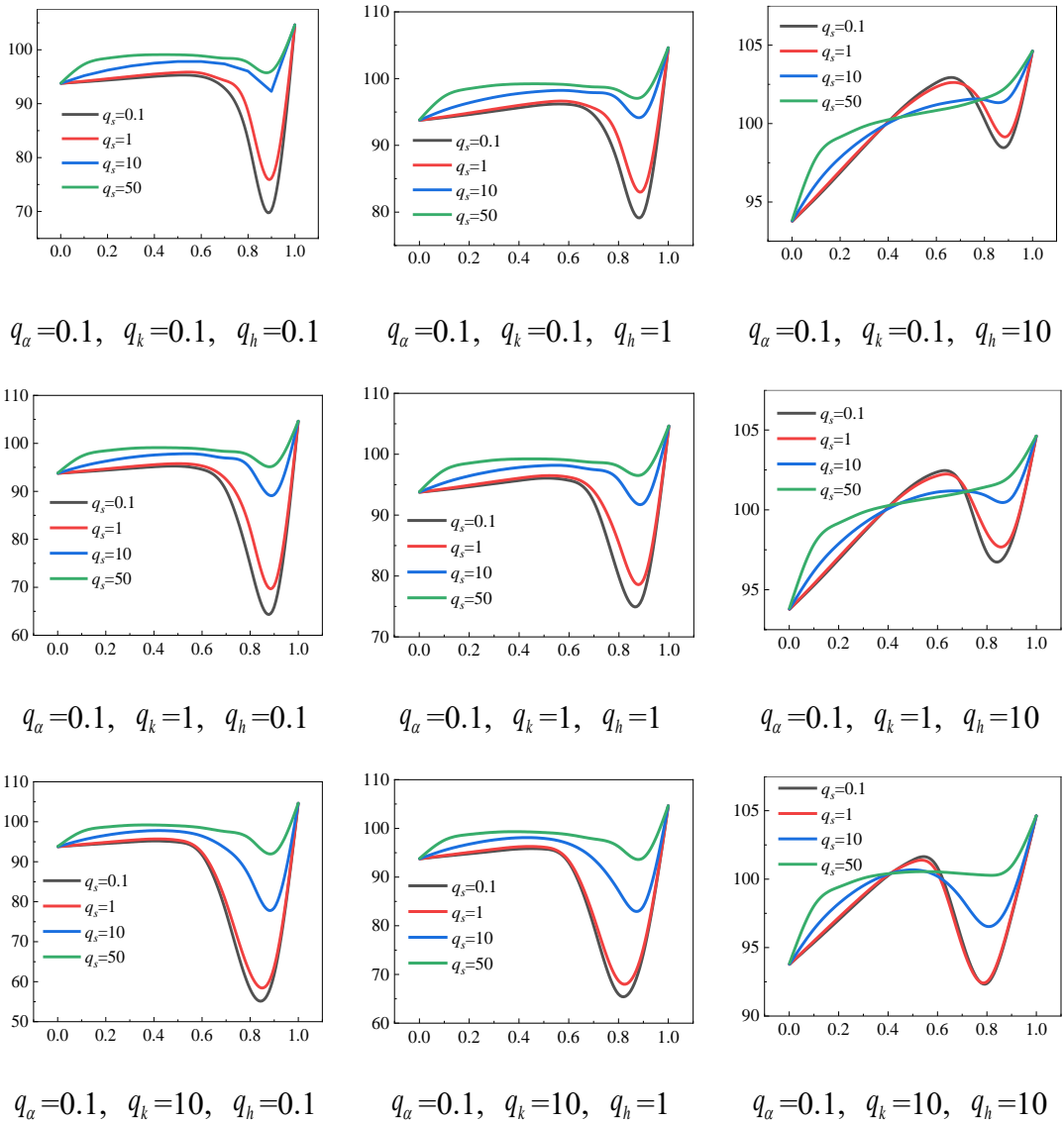

$q_\alpha=0.1$, $q_k=0.1$, $q_h=0.1$     $q_\alpha=0.1$, $q_k=0.1$, $q_h=1$     $q_\alpha=0.1$, $q_k=0.1$, $q_h=10$

$q_\alpha=0.1$, $q_k=1$, $q_h=0.1$     $q_\alpha=0.1$, $q_k=1$, $q_h=1$     $q_\alpha=0.1$, $q_k=1$, $q_h=10$

$q_\alpha=0.1$, $q_k=10$, $q_h=0.1$     $q_\alpha=0.1$, $q_k=10$, $q_h=1$     $q_\alpha=0.1$, $q_k=10$, $q_h=10$

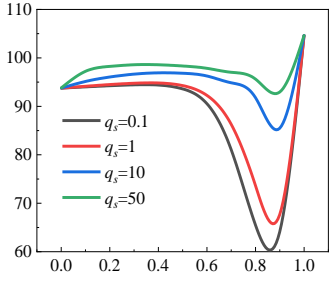

$q_\alpha=1$, $q_k=0.1$, $q_h=0.1$

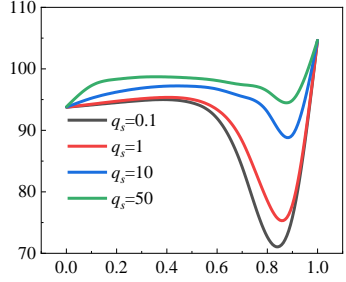

$q_\alpha=1$, $q_k=0.1$, $q_h=1$

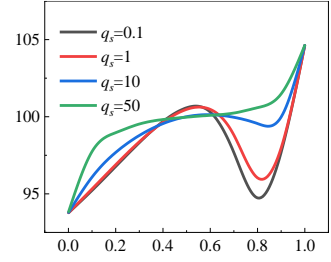

$q_\alpha=1$, $q_k=0.1$, $q_h=10$

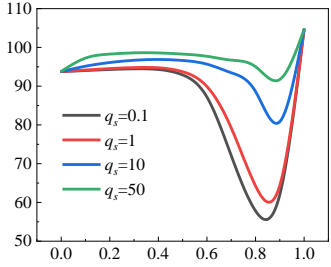

$q_\alpha=1$, $q_k=1$, $q_h=0.1$

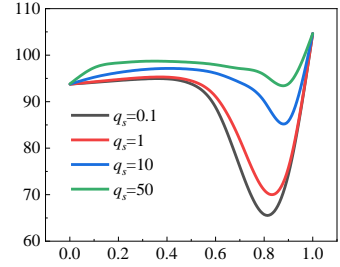

$q_\alpha=1$, $q_k=1$, $q_h=1$

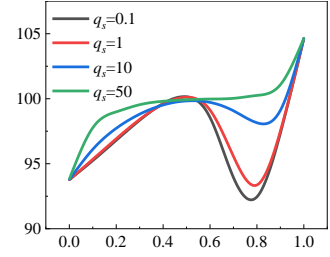

$q_\alpha=1$, $q_k=1$, $q_h=10$

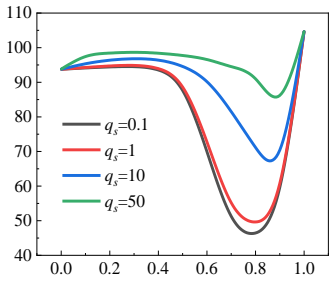

$q_\alpha=1$, $q_k=10$, $q_h=0.1$

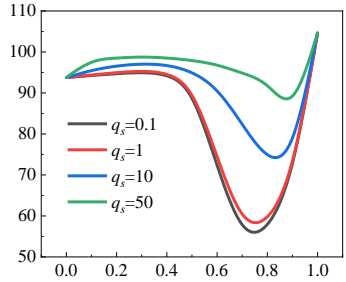

$q_\alpha=1$, $q_k=10$, $q_h=1$

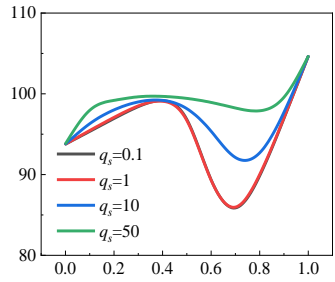

$q_\alpha=1$, $q_k=10$, $q_h=10$

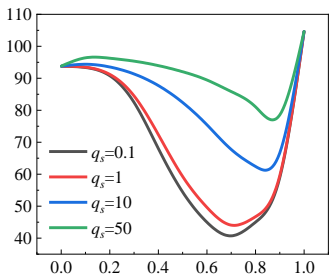

$q_\alpha=10$, $q_k=0.1$, $q_h=0.1$

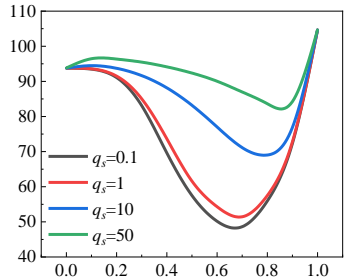

$q_\alpha=10$, $q_k=0.1$, $q_h=1$

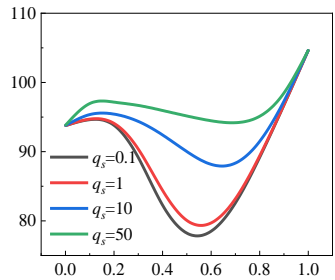

$q_\alpha=10$, $q_k=0.1$, $q_h=10$

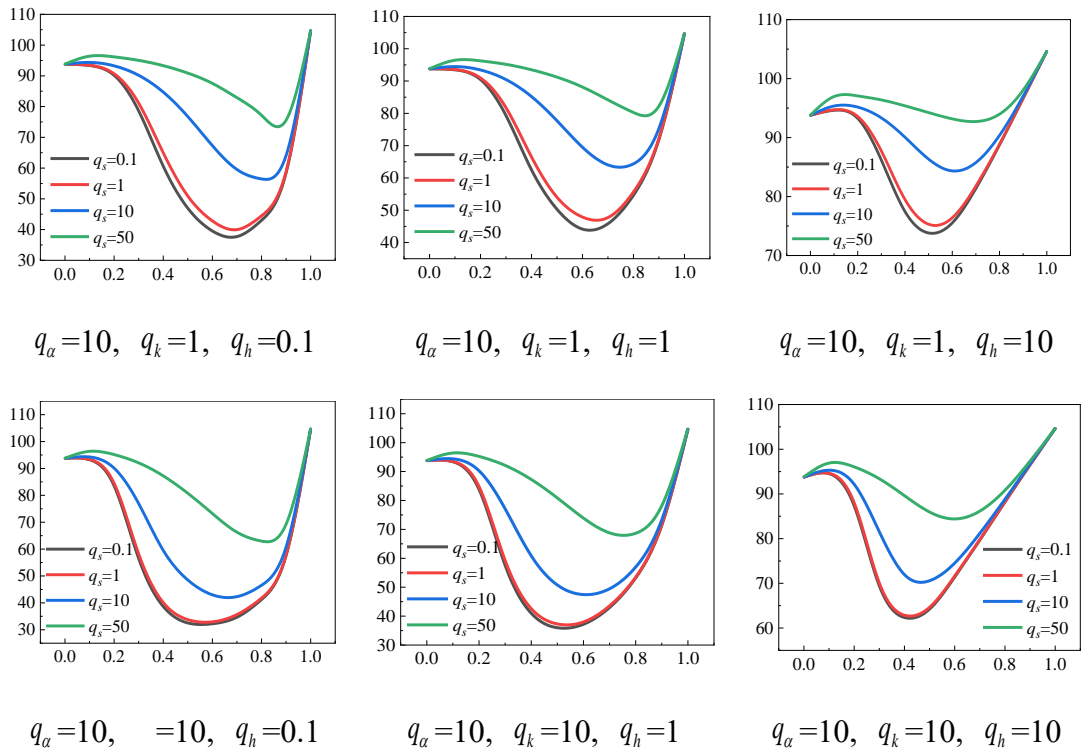

Fig. 10. 27 subgraphs with 108 curves of objective function varies with respect to design variable with 108 kinds of penalty factor combinations.

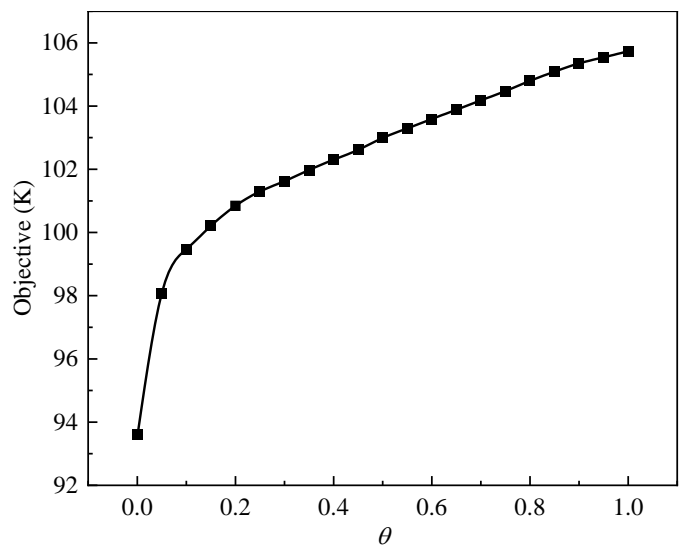

Fig. 11. Objective function varies with respect to the design variable with penalty factors $q_\alpha=0.1$, $q_k=0.1$, $q_h=50$, $q_s=100$.

4.4. Implementation of topology optimization

After the values of penalty factors are chosen, the topology optimization is implemented. In this work, the optimization problem is handled as a constrained minimization problem:

$$\begin{aligned}
\min_{\theta}: \quad & f(\psi(\theta,t),\theta,t) \\
\text{s.t.}: \quad & \gamma(\psi(\theta,t),\theta,t)=0 \\
& \iint_{\Omega_d}(1-\theta)dxdy - f_V A_{\Omega_d} \leq 0 \\
& 0 < \theta(\mathbf{x},t) \leq 1 \quad \forall \mathbf{x} \in \Omega_d, t \in [0,t_T]
\end{aligned} \quad (31)$$

where $\psi(\theta,t)$ represents the field variable vector including the temperature field, velocity field and pressure field; $\gamma(\psi(\theta,t),\theta,t)$ denotes the residual of the finite element formulation of the full thermofluid problem; $\Omega_d$ and $A_{\Omega_d}$ are the design domain and area of the design domain, respectively; $f_V$ is the volume fraction of the constraint set in this model; and $\mathbf{x}$ denotes the spatial coordination vector of the model.

During the process of topology optimization, design filtering is necessary in the transient thermofluid model to avoid checkerboard problems [47]. A partial differential equation (PDE) filter is used in the topology optimization procedure, which defined as:

$$-r_{fil}^2 \nabla^2 \tilde{\theta} + \tilde{\theta} = \theta \quad \text{in } \Omega_d \quad (32)$$

where $r_{fil}$ and $\tilde{\theta}$ represent the filter parameter and filtered design variable, respectively. Because the surroundings are fluid domain, the boundary conditions of the filter PDE can be expressed as follows:

$$\tilde{\theta} = 1 \quad \text{on } \partial\Omega_d \quad (33)$$

To reduce the intermediate density elements on the interface between solid and fluid the design domain, a smoothed Heaviside projection is applied on the filtered design field:

$$\bar{\bar{\theta}} = \frac{\tanh(\beta\eta) + \tanh\left(\beta\left(\tilde{\theta} - \eta\right)\right)}{\tanh(\beta\eta) + \tanh(\beta(1-\eta))} \tag{34}$$

where $\bar{\bar{\theta}}$ denotes the projected design field, $\beta$ is a parameter controlling the slope of the projection function, and $\eta$ is the projection threshold parameter. The projected design field $\bar{\bar{\theta}}$ is substituted with the initial design field in the modified governing Eqs. (21), (23) and (25), and the interpolation functions (26). Sensitivities are subsequently corrected using the chain rule.

Topology optimization of the transient pseudo-3D model is implemented in the commercial FEA software COMSOL Multiphysics. The governing equations (21) and (23) are calculated in the "Heat transfer in solids and fluids" module, coupling with the "Laminar flow" module to obtain the temperature field $T$, the velocity field $\mathbf{u}$ and pressure field $p$. Eq. (25) is implemented in a heat partial differential equation to obtain the temperature field of the base plate $T_{bp}$. The coupled models are solved using the time-dependent FE solver. The optimization method used in this model is GCMMA, with the number of inner iteration per outer step is set to 1.

## 5. Results and discussion

Due to the highly non-linear and non-convex properties of the optimization problem, the optimizer will always converge to a local optimal structure. One method to alleviate this problem is choosing different initial designs of the model. Then, the optimized results are compared with the corresponding full 3D model to validate their performance. Furthermore, steady-state optimized results are also compared to highlight the characteristics of the transient model. Lastly, a practical model is built with more powerful cooling medium and higher inlet pressure.

5.1. Initial design

In this section, several initial designs are proposed and then an overall comparison is made to select a better initial design for the topology optimization. Four different initial design layouts are shown in Fig. 12. The first represents a uniform design field ($\theta_0$ =0.8) in the design domain and the other figures are three solid cylinders in different positions of the design domain. It is necessary to figure out which initial design has a relatively better convergence result. The input pressure for the topology optimization procedure is set to 1Pa rather than 2Pa, because the transient solver cannot converge under the prescribed tolerance during optimization in the software. As the inlet condition changes, the heat exchange coefficient correspondingly changes to $h_s = 6.5 \times 10^4 \text{W/(m}^2 \cdot \text{K)}$ and $h_f = 90 \text{W/(m}^2 \cdot \text{K)}$. The other optimization parameters are shown in **Table 3**.

The choice of 4 initial designs in Fig. 12 relies on several reasons: Firstly, the comparison of uniform and non-uniform density distribution need to be made; Secondly, because the design domain and the model is axisymmetric, the density layouts selected in this section are axisymmetric as well; Lastly, the layout of rounds that represent cylinder heat sink fins in full 3D model, in Initial design 2-4 refers to the layout of initial designs of *model a* in [48] to compare the performance of different layout directions and fins number.

In order to reduce the influence of finite element analysis accuracy on the optimization result, local mesh refinement is shown in Fig. 13. Symmetry is adopted in the optimization process in order to reduce computational effort.

After optimization, 4 initial designs converge to the optimized structures shown in Fig. 15, whose physical fields are shown in Fig. 16, Fig. 17 and Fig. 18. The objective convergence curves are shown in Fig. 14 and performance measures of the 4 optimized designs and a reference are shown in Fig. 19. In Fig. 19, only the optimized 1, 2, and 4 and two regular heat sinks are displayed because the performance curve of $3^{th}$ optimized is too similar to the $2^{th}$ to separate them. Besides, the "*V*" displayed in Fig. 19 represents the total volume of the design domain with a value of 640mm$^3$.

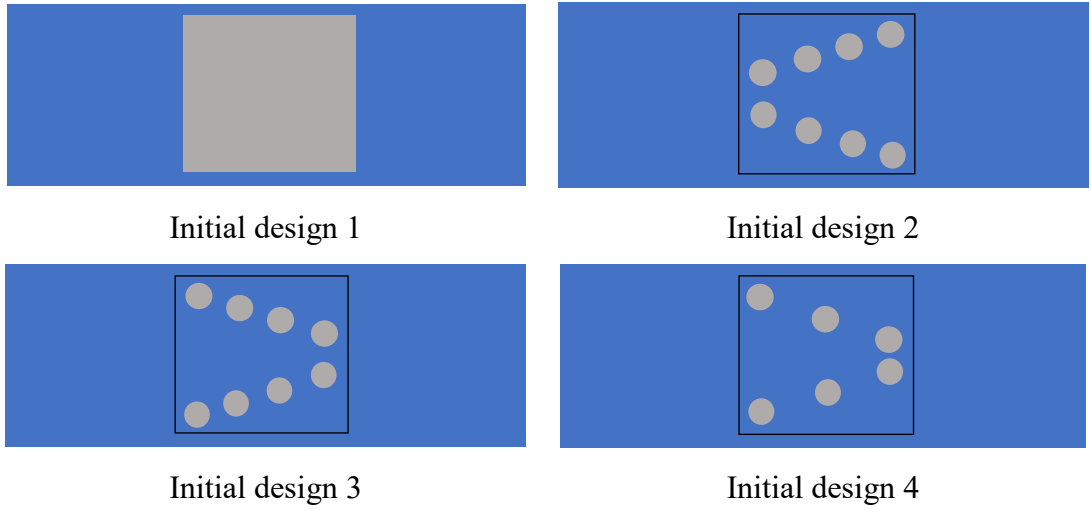

Initial design 1　　　　　　　　　　　　Initial design 2

Initial design 3　　　　　　　　　　　　Initial design 4

Fig. 12. Four different initial designs of the pseudo 3D topology model.

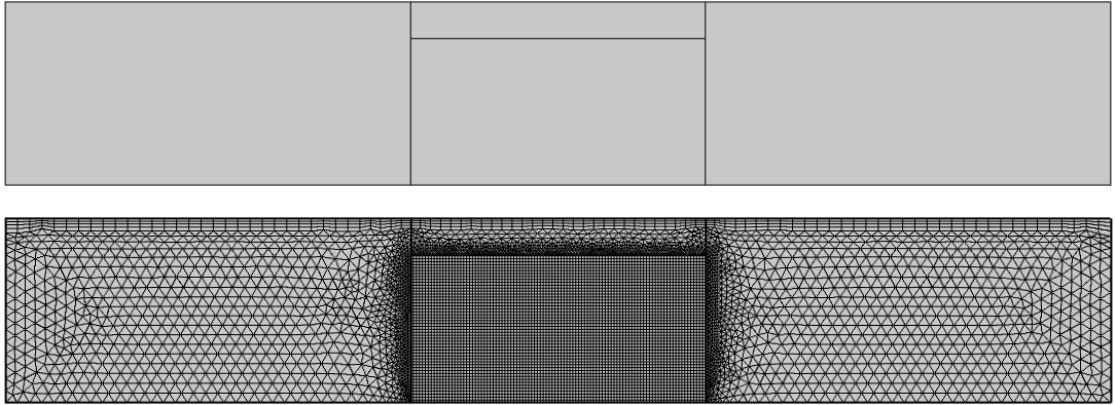

Fig. 13. The sketch of topology optimization model and the mesh.

**Table 3**

Parameters of optimization process.

| Parameters of optimization | Values | Parameters of optimization | Values |
|---|---|---|---|
| $Da$ | $1\times10^{-6}$ | $q_\alpha$ | 0.1 |
| $n_{ev}$ | 300 | $q_k$ | 0.1 |
| $\varepsilon$ | $1\times10^{-3}$ | $q_h$ | 50 |
| $d_{min}$ [mm] | 0.133 | $q_s$ | 100 |
| $r_{min}$ [mm] | 0.2 | $\beta$ | 8 |
| $p_{in}$ [Pa] | 1 | $\eta$ | 0.5 |
| $f_v$ | 0.5 | | |

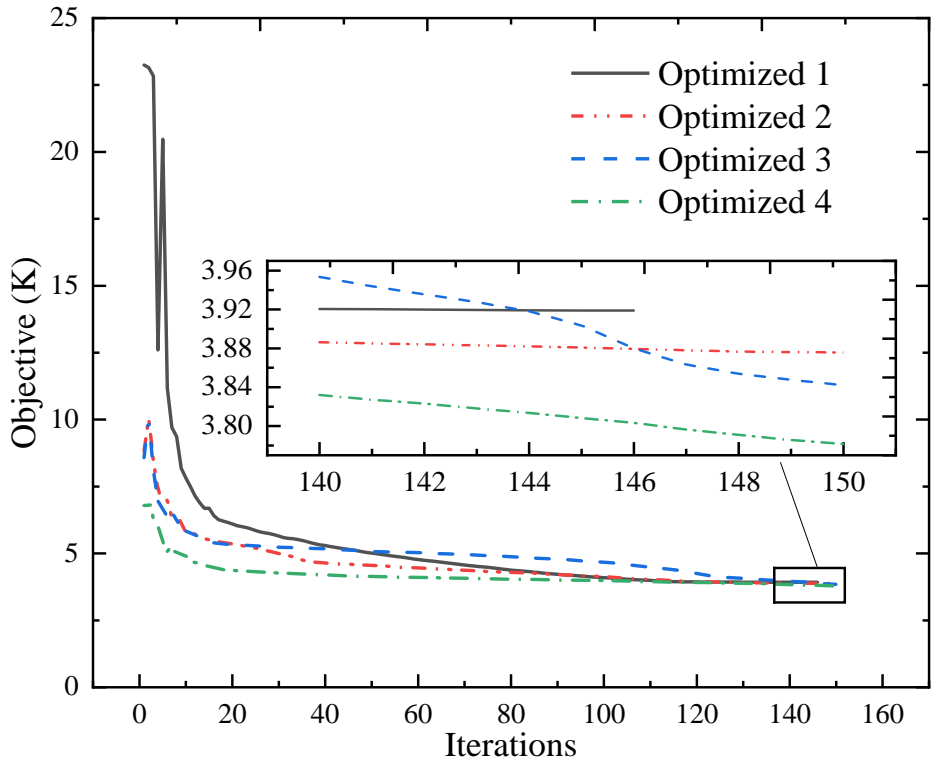

Fig. 14. Objective function with respect to the iteration numbers of initial designs 1, 2, 3 and 4.

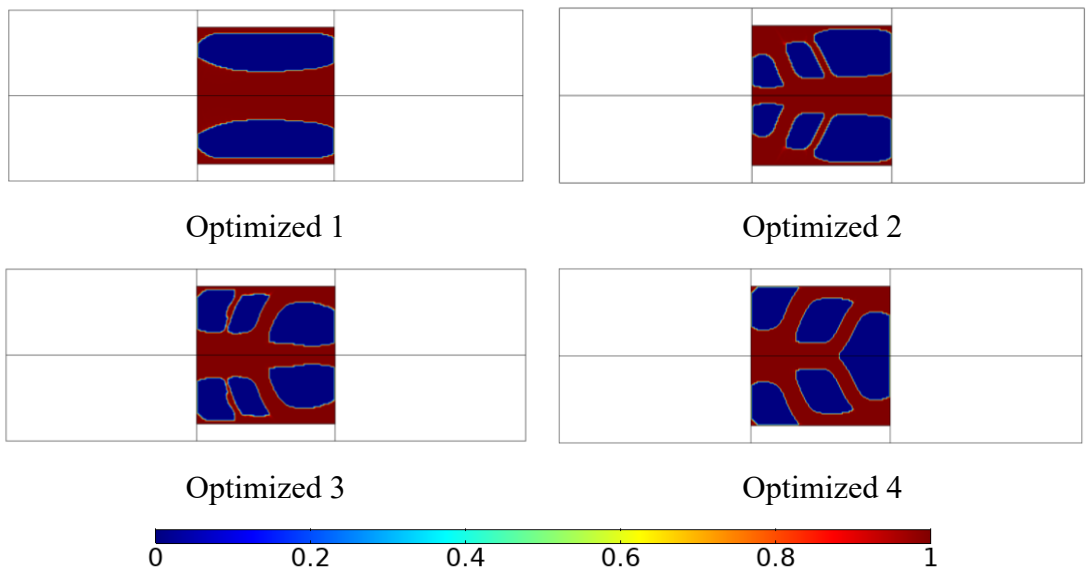

Fig. 15. Four optimized structures obtained from four different initial designs.

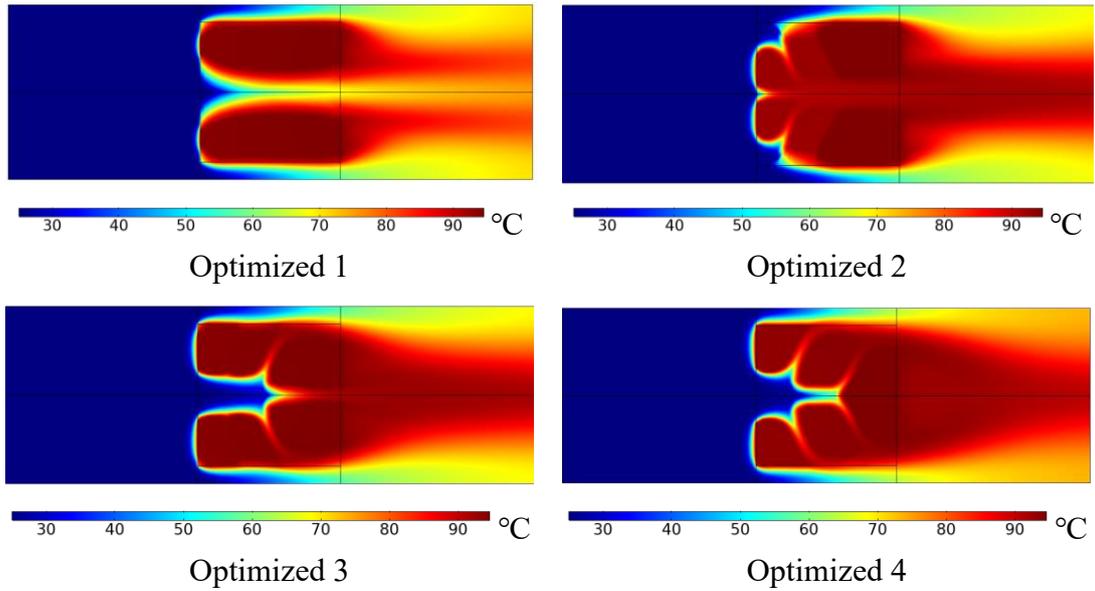
Fig. 16. Temperature distribution of the channel layer for the 4 optimized structure.

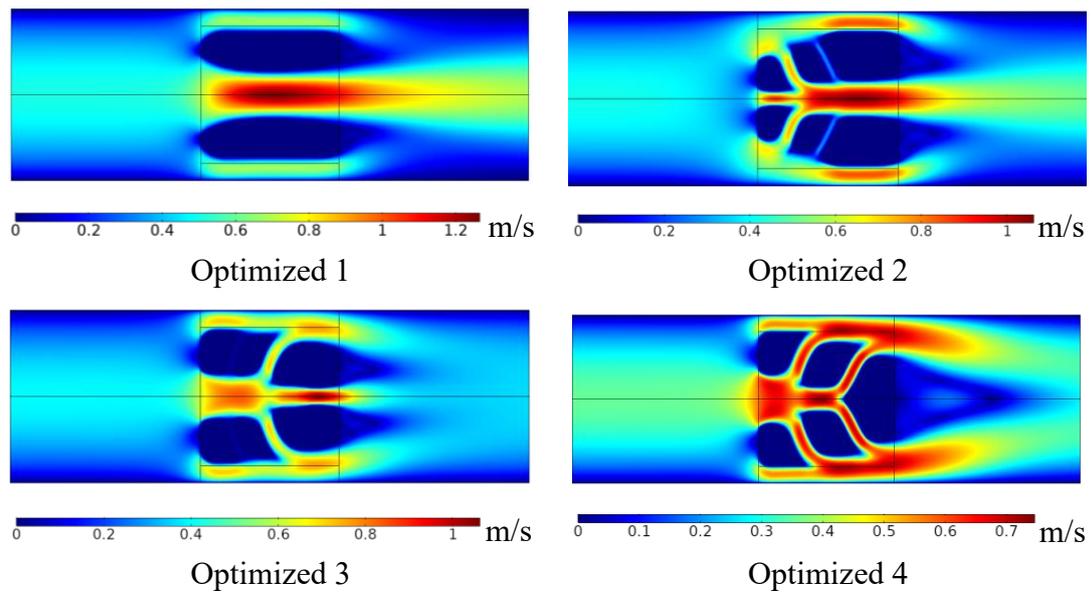
Fig. 17. Velocity distribution of the channel layer for the 4 optimized structure.

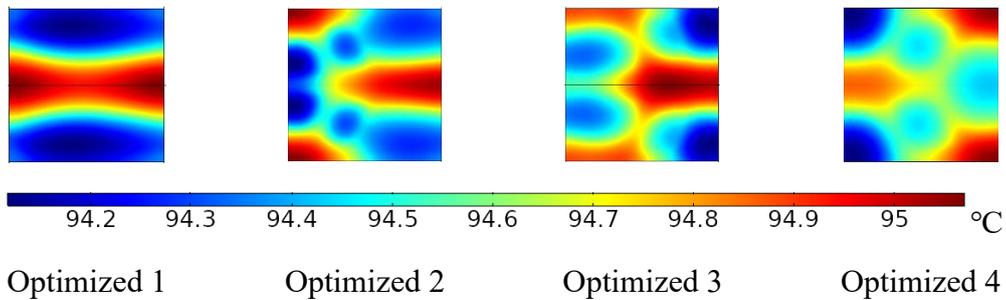
Fig. 18. Temperature distribution of the base plate layer for the 4 optimized structure.

According to Fig. 14, the 4$^{th}$ initial design produces the best performing design with respect to minimum average baseplate temperature. In Fig. 15, it can be found that the fluid channels through the design are generally thicker, than for the other designs. This corresponds to higher fluid velocities through the heat sink geometry, as shown in Fig. 17. This translates to overall higher heat dissipation through convection and a better cooling of the base plate layer. It is interesting to note that in Fig. 16, the 2$^{nd}$ initial design has lower temperatures in the foremost fins than the other designs. This only translates to lower local base plate temperatures in Fig. 18, but as seen from Fig. 14, a slightly higher average base plate temperature than the 4$^{th}$ initial design.

The transient pseudo-3D model is an approximate model of the full 3D situation. Furthermore, the design layout is determined by the utility of Brinkman penalization, in which the physical field boundaries are not as clear as for a pure fluid and solid structure. By setting a threshold of the design field at a predefined value of 0.8, this allows for extrusion of the design to 3D for verification.

For comparing the pseudo-3D and full 3D models, we chose two performance measures:

a) The average temperature $T_{avg}$ of the heat source;
b) The pumping power $P_{pump}$ required to cool heat sink.

The consuming pumping power of the heat sink is defined as follows:

$$P_{pump} = r_f \cdot p_{drop} \qquad (35)$$

where $r_f$ represents the volumetric flow rate passing through inlet and $p_{drop}$ is the pressure drop of the overall heat sink. The volumetric flow rate $r_f$ is formulated as:

$$r_f = v_{in} A_{in} \qquad (36)$$

Lastly, the velocity at the inlet $v_{in}|_{t=1s}$ [m/s] and the heat sink volume $V_{hs}$ are important to compare the various designs.

**Table 4**

Performance measures of 4 different optimized result and regular structure.

| Structure | $T_{avg}|_{t=1s}$ [°C] | $v_{in}|_{t=1s}$ [°C] | $V_{hs}$ [mm³] | $P_{pump}|_{t=1s}$ [mW] |
| --- | --- | --- | --- | --- |
| Reference(0.375$V$) | 94.97 | 0.369 | 240 | 0.0369 |
| Optimized 1 | 94.72 | 0.524 | 320 | 0.0524 |
| Optimized 2 | 94.54 | 0.315 | 320 | 0.0315 |
| Optimized 3 | 94.53 | 0.314 | 320 | 0.0314 |
| Optimized 4 | 94.24 | 0.242 | 320 | 0.0242 |
| Reference(0.5$V$) | 94.38 | 0.257 | 320 | 0.0257 |

As shown in Table 4, the average temperature of the heat source at 1s of the optimized 4 design is 0.721°C lower than that of reference (0.375$V$) design. In terms of pumping power, the optimized 4 design decreases the pumping power by 34.3% of that for the reference heat sink (0.375$V$). This indicates that the optimized design only uses 65.7% of the pumping power and reduced the volume average temperature of the chip by 0.721°C at $t$ =1s, both compared to the reference design. As for the comparison with the reference heat sink (0.5$V$), the optimized heat sink decrease only 0.14°C of temperature and 5.1% of pumping power.

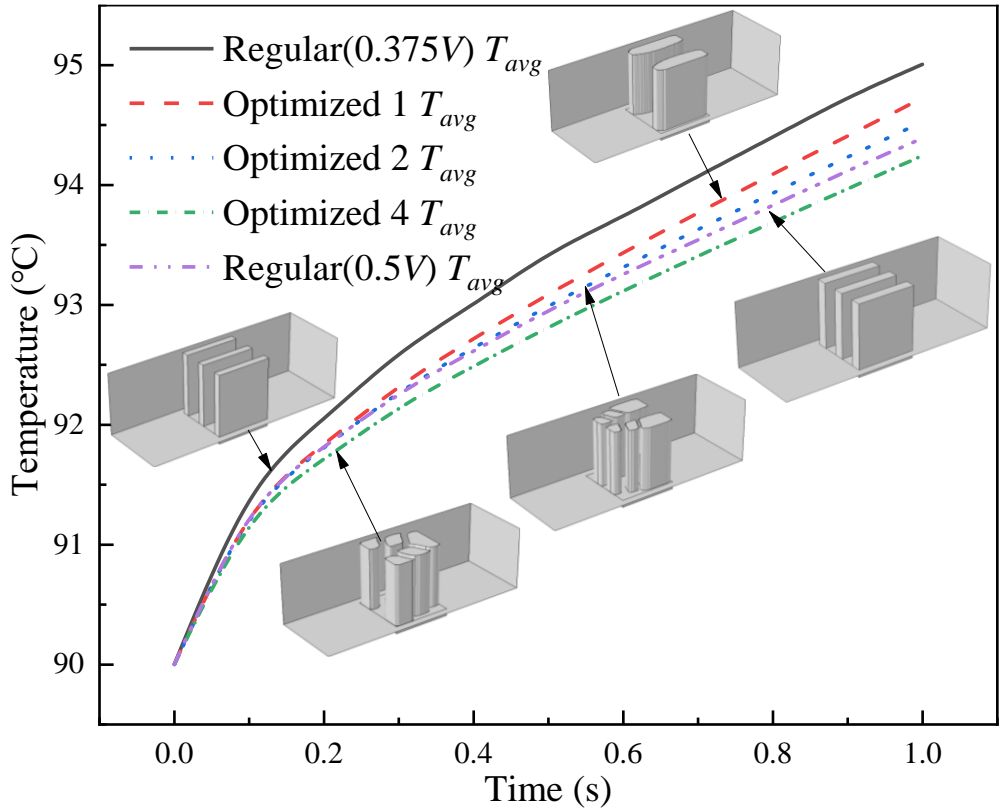

Fig. 19. Average temperature of chip of different structures

5.2. Comparison with steady-state pseudo topology optimization model

To highlight the importance of using a transient model for treating the instantaneous behavior, the transient pseudo-3D model is compared to a steady-state pseudo-3D model. The comparison is divided into two parts:

a) The superior steady-state heat dissipation performance of the steady-state optimized structure compared to the reference design;
b) The superior instantaneous thermal performance of the transient optimized structure compared to the steady-state optimized structure.

To make it comparative, the initial design of the steady-state optimization is the same as the Initial design 4 in Fig. 12, which performs best out of the four different initial designs for the transient case.

The mesh of the steady-state pseudo 3D model is the same as that of the transient model as shown in Fig. 13. The constraint that represents the volume fraction of the entire design domain of the optimization are set to $f_V = 0.5$. The objective function of the optimization is set to the average temperature of the chip when it reaches steady state. The boundaries of solid and fluid phase chosen for the thresholded surface is a design variable of 0.8 too. Therefore, according to the thresholded boundaries, the geometry is extruded to the 3D model to verify its thermal performance.

As is illustrated in Fig. 21 and Fig. 22, the steady-state average chip temperature of the steady-state optimized structure is about 20°C lower than that of the reference design. This may be because steady state design has much sharper features and more fins than the reference one. This validates the superior steady-state thermal performance of the design generated by topology optimization.

For comparison, the transient performance of the steady-state design is now compared to the transient pseudo-3D topology optimization model. The average temperature is displayed in Fig. 23 and Fig. 24, for the time periods [0, 1s] and [0, 200s], respectively. As can be seen in Fig. 23, the instantaneous thermal performance of the transient pseudo-3D model is better than that of steady-state pseudo 3D model. However, as time goes on above 8 seconds, the steady-state design becomes the better performer. This clearly shows that if the instantaneous transient response is of importance, steady-state analysis is not good enough on the condition in this section.

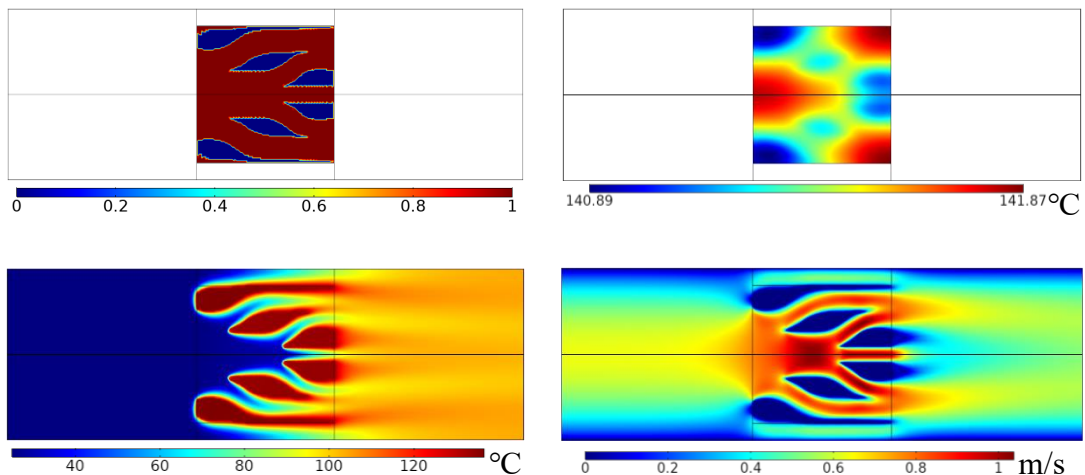

Fig. 20. Design variable layout, base plate temperature, channel temperature and channel velocity of steady-state topology optimized pseudo 3D model.

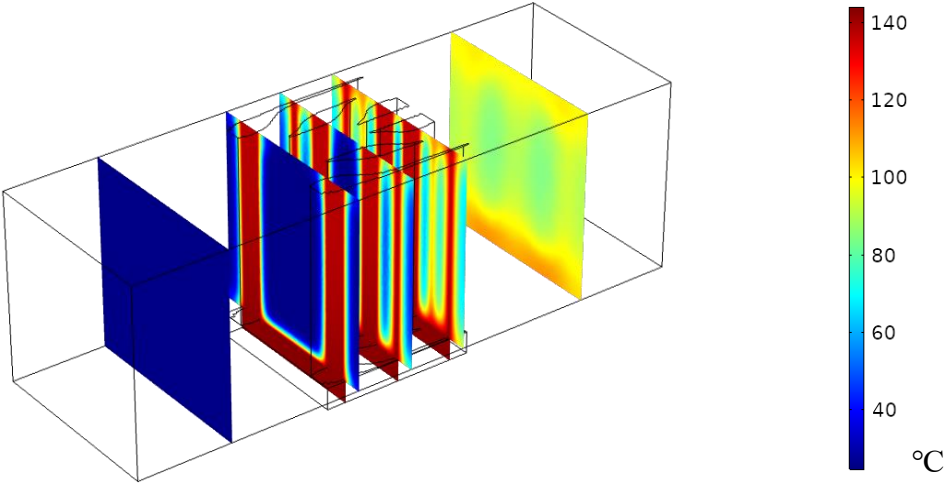

Fig. 21. Temperature layout of optimized steady-state 3D model.

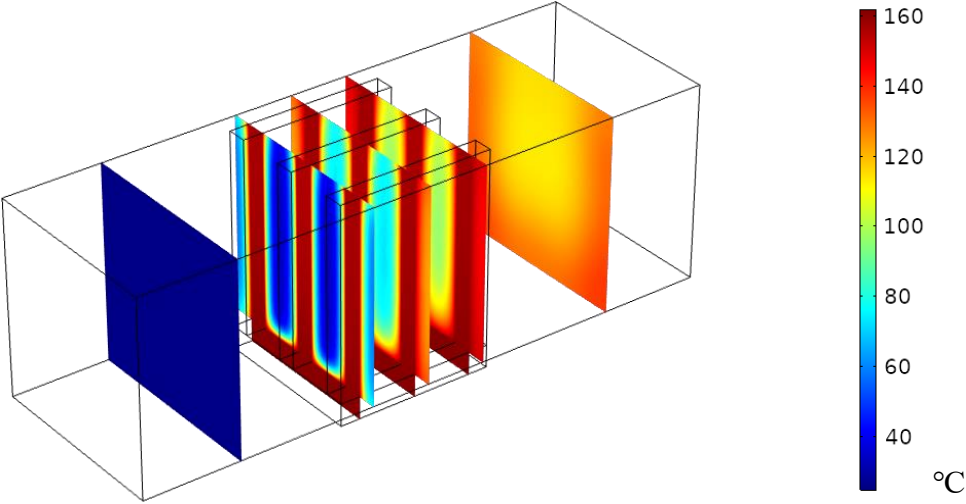

Fig. 22. Temperature layout of reference 3D model.

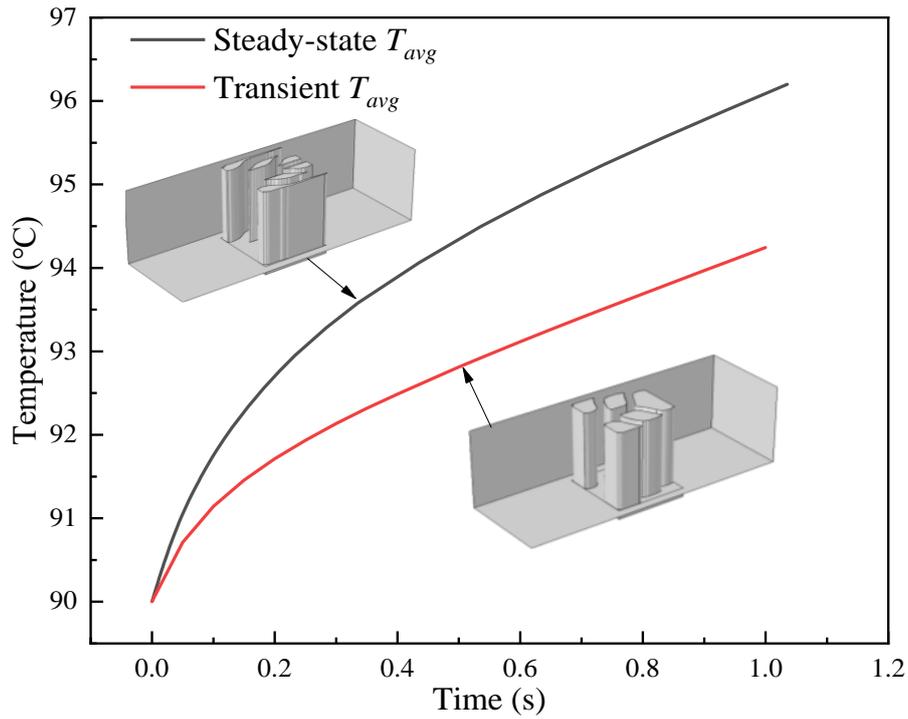

Fig. 23. The average temperature of optimized steady-state 3D model and optimized transient 3D model during [0, 1s].

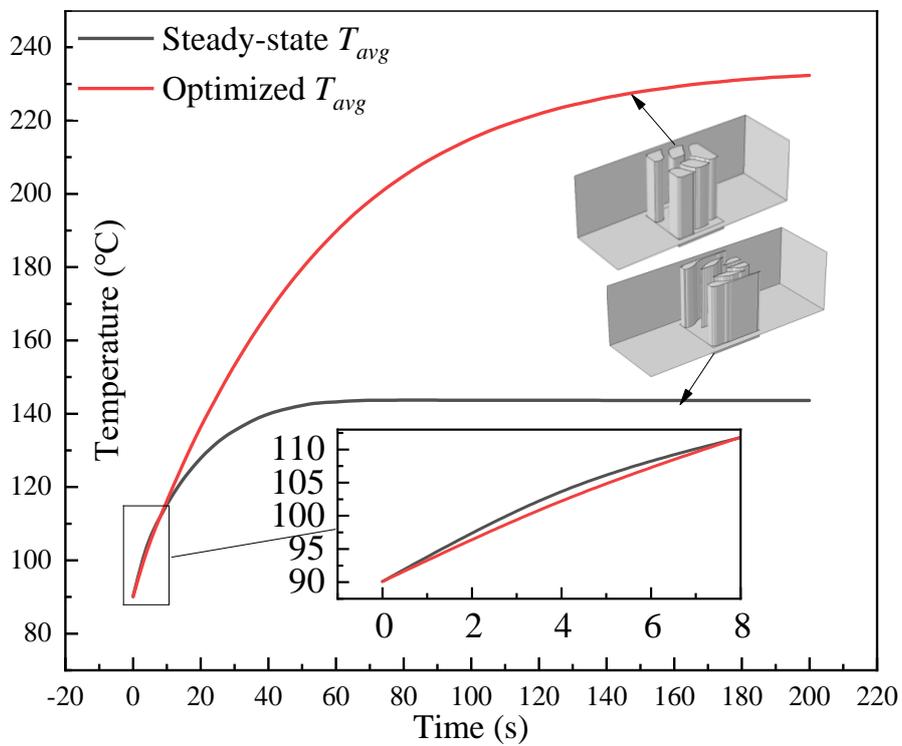

Fig. 24. The average temperature of optimized steady-state 3D model and optimized transient 3D model during [0, 200*s*].

### 5.3. A practical problem

To apply the proposed model to a more practical situation, a more powerful cooling medium and more realistic pressure input than the previous model are adopted in this section. Thus, water is chosen as the cooling medium to replace air and the pressure input rises to 50Pa. Since the thermal conductivity and the specific heat are much higher than that of air, a shorter time interval with terminal time $t_T = 0.1$s is chosen. The solid phase of the model is still aluminum.

The length of the channel is reduced to 12*mm*, which aims to reduce the degrees-of-freedom of the non-design domain in order to cut down the calculation time and allow a finer mesh in the design domain. The 3D model and its corresponding pseudo-3D model are shown in Fig. 25.

As the material properties and the boundary condition change, the heat transfer coefficients $h_s$ and $h_f$ are updated too. The transient 3D model shown in Fig. 1 is use to obtain the value of $h_s$ and $h_f$, according to the Eqs. (15)-(19). Therefore, the value of material properties and heat transfer coefficients are displayed in **Table 5** and **Table 6**. The temperature profiles of the base plate for the pseudo-3D and full 3D models are demonstrated in Fig. 26. The similarity criterion $g$ in Expression (22) is used to compare the temperature profiles, with a value of $g = 5.62 \times 10^{-3}$ showing a high similarity of the base plate temperature profile for the two models.

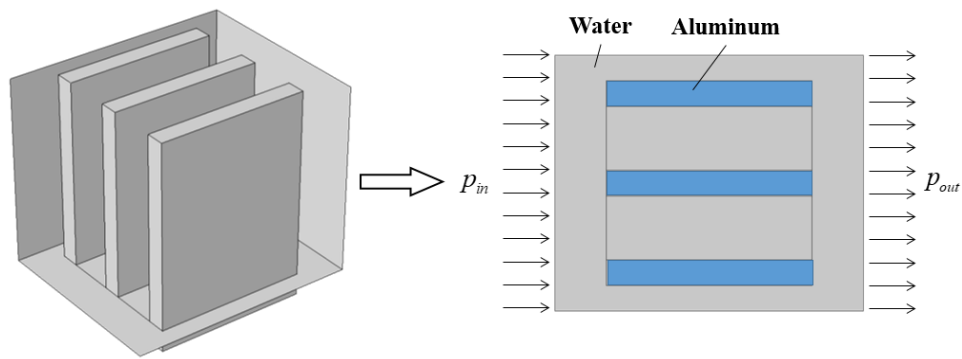

Fig. 25. Geometric diagram of 3D model and the corresponding pseudo-3D model.

**Table 5**

Parameters of boundary conditions and model property of pseudo 3D model and 3D model.

| Parameters | Pseudo 3D model | 3D model |
|---|---|---|
| $T_{in}$ [℃] | 25 | 25 |
| $t_{ch}$ [mm] | 10 | 10 |
| $t_{bp}$ [mm] | 1.15 | 1.15 |
| $Q$ [W] | 50 | 50 |
| $p_{in}$ [Pa] | 50 | 50 |
| $p_{out}$ [Pa] | 0 | 0 |
| $h_s$ [W/(m²·K)] | 2×10⁵ | — |
| $h_f$ [W/m²·K] | 5×10³ | — |

**Table 6**

Thermo-physical properties of pseudo 3D model.

| Thermo-physical properties | Values |
|---|---|
| $k$ [W/(m·K)] | 0.6 |
| $k_s$ [W/(m·K)] | 237 |
| $\rho_f$ [kg/m³] | 988 |
| $\rho_s$ [kg/m³] | 2,700 |
| $c_f$ [J/(kg·K)] | 4,185 |
| $c_s$ [J/(kg·K)] | 900 |
| $\mu$ [Pa·s] | 8.94×10⁻⁴ |

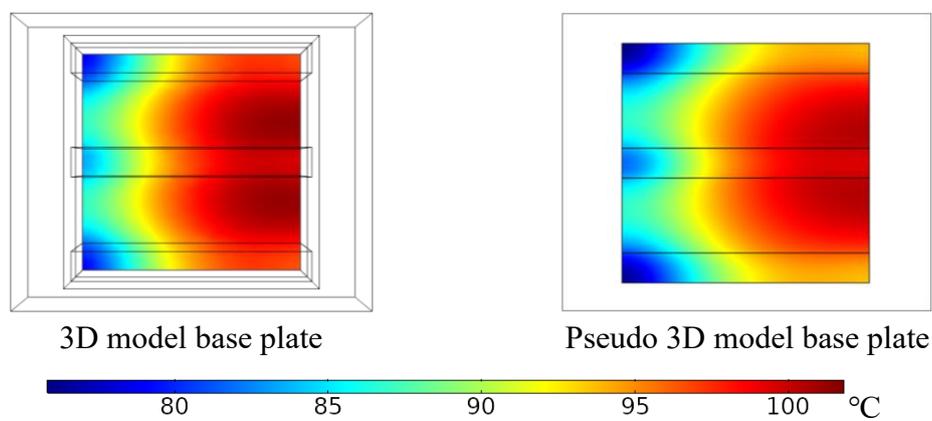

Fig. 26. The temperature profiles of base plate of pseudo 3D model and 3D model.

Besides, the penalty factors $q_\alpha = 0.1$, $q_k = 0.1$, $q_h = 0.1$ and $q_s = 100$ are chosen in this section. The reason why the $q_h$ selected in this section is different from the previous one is that different $h_s$ and $h_f$ are used here. The objective function curve with respect to design variable $\theta$ is shown in Fig. 27. A monotonously increasing curve should yield a relatively low proportion of intermediate design variables in the final topology optimization result.

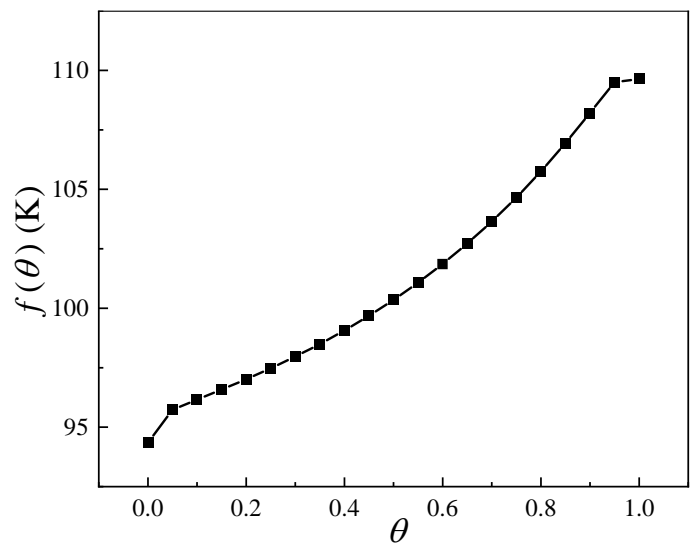

Fig. 27. Objective function curve with respect to design variable.

The sketch and mesh of the model are shown in Fig. 28, in which a very fine quadrilateral mesh is used, as well as boundary layer regions. Fig. 29 shows best performing initial design variable layout from Section 5.1, which is also used in this study.

The optimization configuration is the same as the optimization procedure in Section 5. After 140 iterations of optimization as shown in Fig. 30. The design variables converges to a layout shown in Fig. 31, along with the channel temperature, channel velocity and base plate temperature fields.

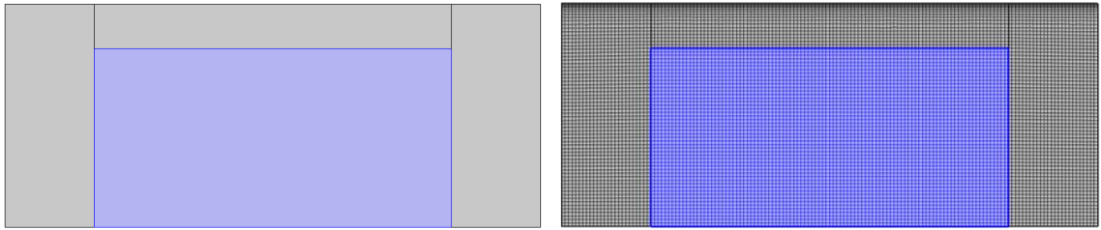

Fig. 28. The sketch of topology optimization model and its meshes.

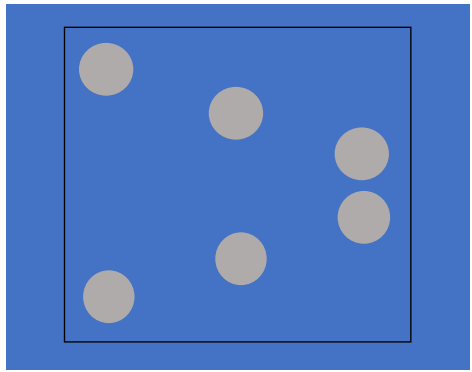

Fig. 29. Initial design of the pseudo 3D topology optimization model.

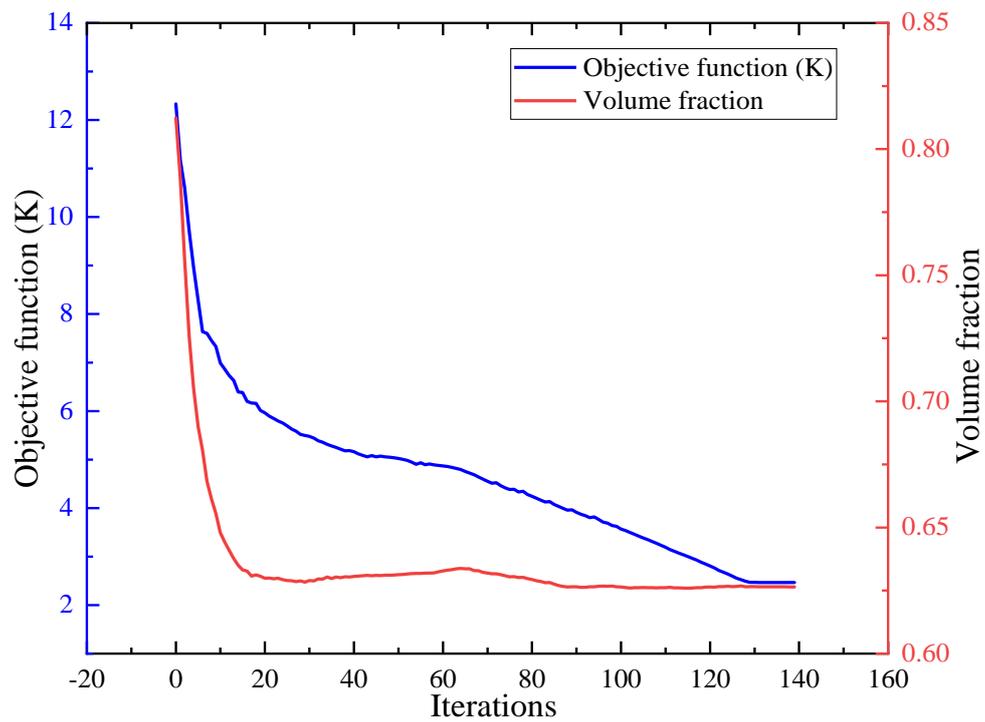

Fig. 30. Objective function and volume fraction of transient topology optimization pseudo 3D model with respect to iteration number.

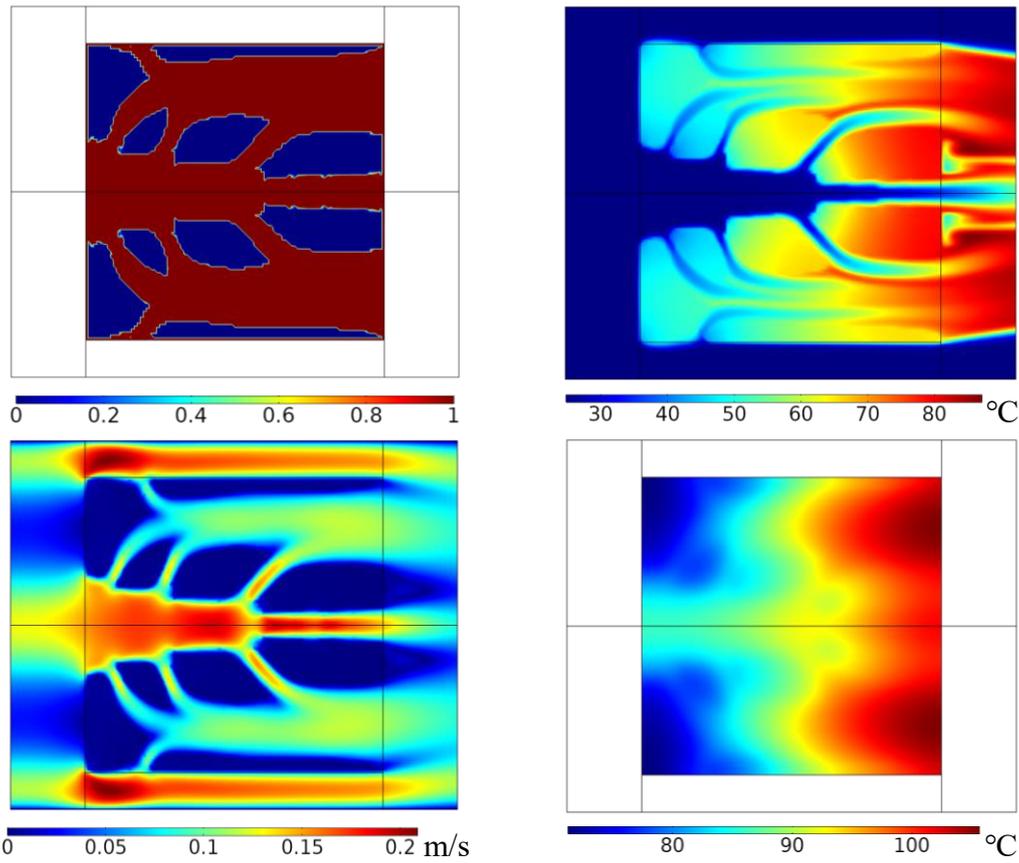

Fig. 31. Design variable layout, channel temperature, channel velocity and base plate temperature of the topology optimized pseudo 3D model.

**Instantaneous comparison**

In order to verify the performance of the optimized design, the threshold design is extruded to form a full 3D model. It is worth mentioning that since the converged volume fraction is 0.625, as shown in Fig. 30, the stretching model has the same solid material consumption as the ordinary model. Therefore, comparisons in terms of average baseplate temperature and pumping power with the reference straight-fin heat sink are made during a period of 0 and 0.1s are made. This can be seen in Fig. 32 and Fig. 33, where it is obvious that the optimized structure shows a more instantaneous performance than the ordinary structure and it consumes a relatively lower pumping power.

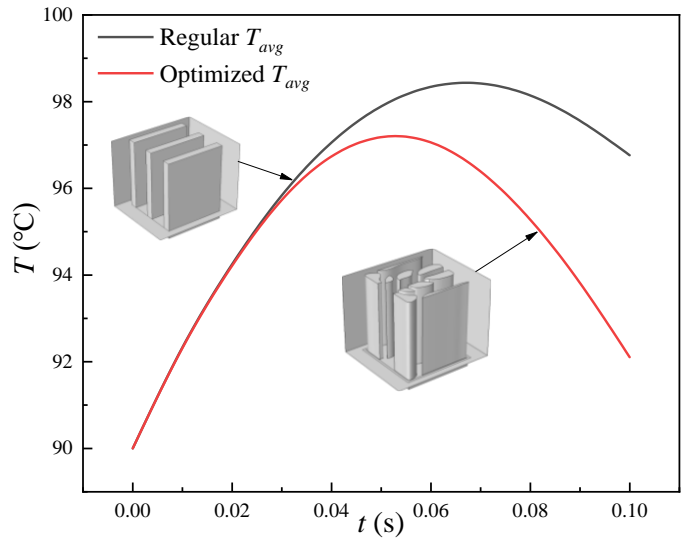

Fig. 32. Average base plate temperature of optimized and reference heat sink.

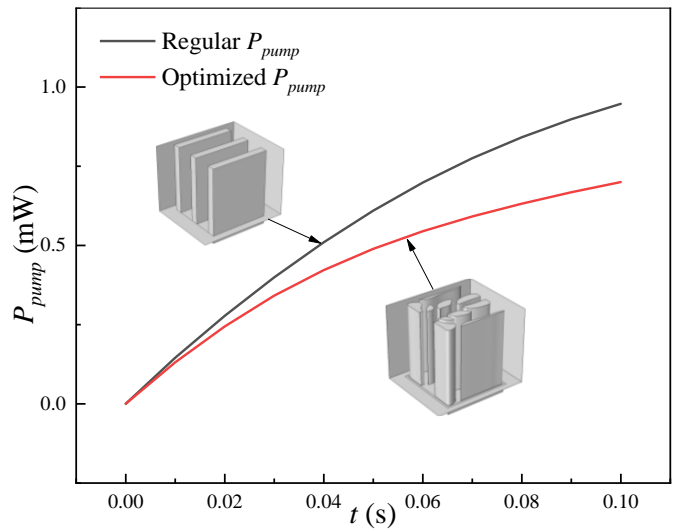

Fig. 33. Pumping power of optimized and reference heat sink.

**Quasi-steady state comparison**

Although the optimized structure performs significantly better instantaneously, while being more energy-efficient, the steady-state performance must also be compared with the reference heat sink. The time interval is extended to 200s, which is used to ensure that models can reach a steady-state. Thus, the average baseplate temperature of the two models are shown in Fig. 34, from which the steady-state temperature of the optimized design is around 5°C lower than the reference design. In comparison to the previous example using air, it is observed that the temperature will decrease below the turbo limit temperature of 90ºC after approximately 0.11 seconds. In Fig. 34, when the average temperature reaches 55°C, which is an assumed lower limit temperature in this section to study their rising temperature procedure, the cooling time period $t_c$ of each model are acquired: $t_{c,opt}$ and $t_{c,reg}$ respectively. In Fig. 35, pumping powers between 0 and 200s of optimized and reference heat sink are compared, in which there are optimized and reference heat sink pumping power converging to 0.964mW and 1.47mW. Therefore, the optimized heat sink can achieve a lower quasi-state temperature with a relative lower pumping power.

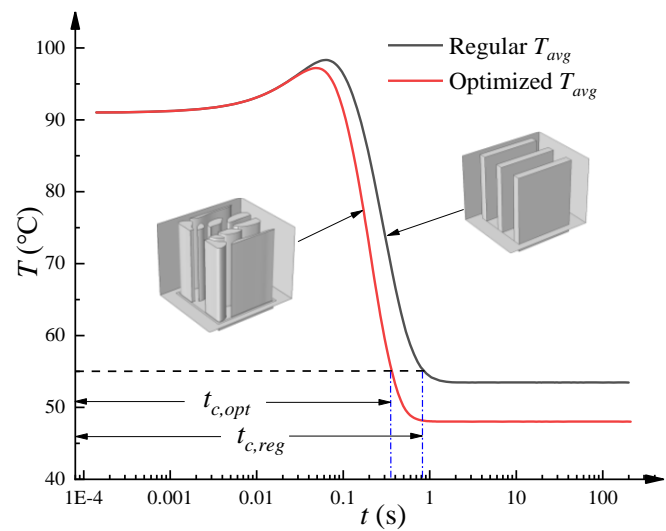

Fig. 34. Average base plate temperature of the optimized and reference heat sink.

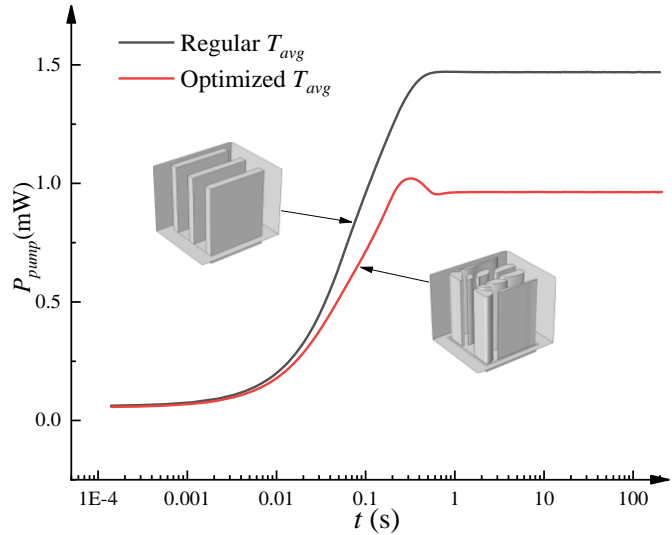

Fig. 35. Pumping power of the optimized and reference heat sink

**Comparison on practical conditions**

As the working conditions above are very ideal and have high energy consumption, a more practical setup is introduced. There are now two critical temperature limits: 90°C and 55°C. When the average temperature of the baseplate, which is usually measured by sensors in the CPU chips, reaches 90°C, the cooling system will activate. During operation, a pressure of 50Pa is imposed on the inlet. With the increasing pumping power, the flow in the channel is accelerated, which leads to a constant decrease in the baseplate temperature. When the temperature then is reduced to 55°C, the cooling system stops. Obviously, the temperature will rise again back to 90°C, and the cooling system will be activated with the whole procedure repeated.

The simulation results for the optimized and reference designs under such working conditions are shown in Fig. 36. It can be seen that the cooling time of the optimized model, $t_{c,opt}$ = 0.36s, is significantly faster compared to the cooling time of the reference design, $t_{c,reg}$ = 0.82s. In terms of heating time, the optimized design has a very close result to the reference design: $t_{ht,opt}$ = 1.55s versus $t_{ht,reg}$ = 1.54s. This is likely because the two designs have similar mass and thus a similar thermal mass.

In terms of pumping power, the total energy that two designs require is calculated with the formulation as follows:

$$E = \int_{t_c} P_{pump} dt \tag{37}$$

where $t_c$ represents the cooling time period. Thus, the pumping energy that the designs consume shown in Fig. 36 can be obtained as: $E_{opt} = 0.28 \times 10^{-3}$ J versus $E_{reg} = 1.04 \times 10^{-3}$ J. The working period of the optimized design lasts 1.913s and the reference design lasts 2.367s. Therefore, the equivalent average energy consumption rate of the two designs are $ES_{opt} = 0.146 \times 10^{-3}$ J/s and $ES_{reg} = 0.439 \times 10^{-3}$ J/s. From this, it can be concluded that the optimized design can save 66.7% of the pumping energy of the reference design over the same time period.

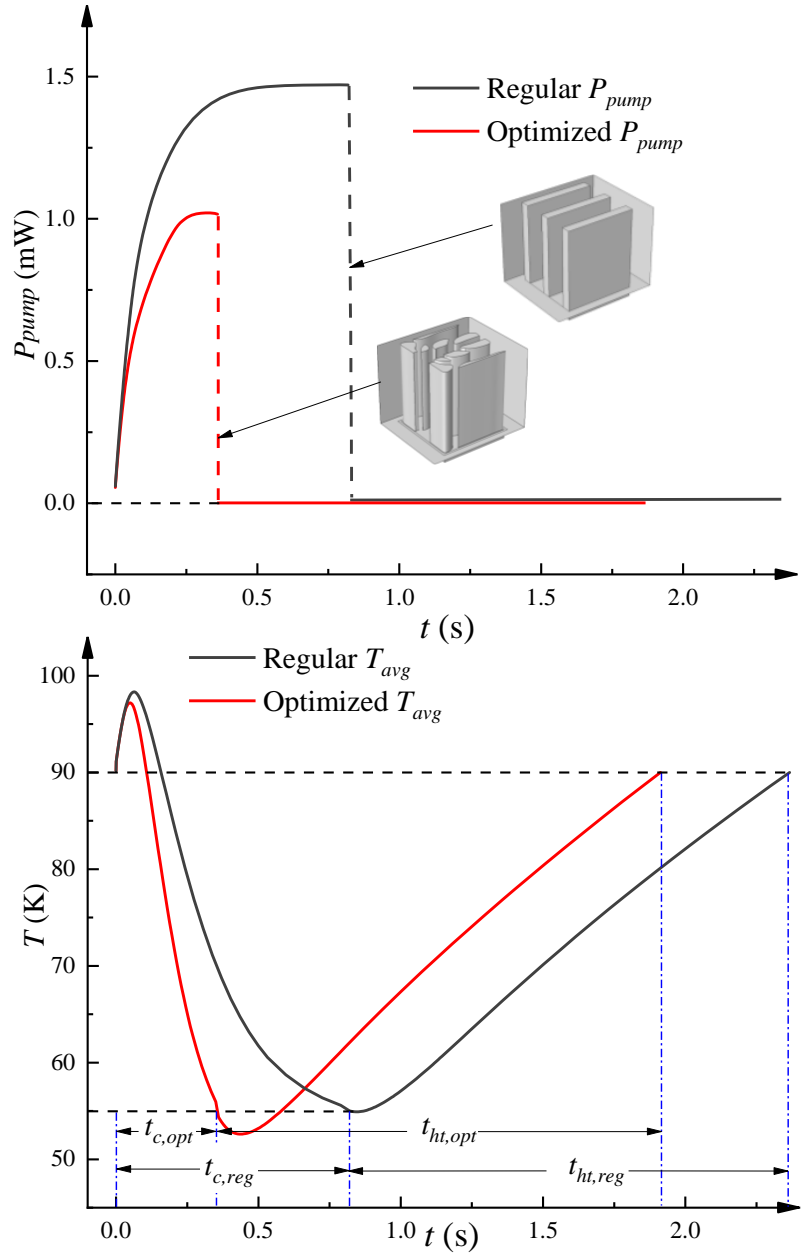

Fig. 36. Performances of two models in a period

## 6. Conclusions

In this work, a transient pseudo-3D simulation and optimization model is introduced to provide a computationally cheap method to fulfill the transient heat flow topology optimization process of 3D models. The optimization of pseudo-3D heat sinks have much lower manufacturing and processing requirements due to their inherent extrudability.

Although the transient pseudo-3D model is an approximate model of full 3D one, the proposed method provides a no more than 1% error of base plate temperature profile according to the introduced criterion. The proposed method mainly relies on the introduction of an artificial heat convection coefficient to establish the approximate relationship between pseudo 3D and full 3D. Thus the value of the heat convection coefficient decides the accuracy of pseudo-3D model. The value of the heat convection coefficient can be obtained by Eqs. (15)-(19) in the full 3D simulation.

Besides, the detailed investigation of interpolation parameters and the choice of monotonously increasing interpolation curves assure a result with less intermediate density.

With the choice of the proper initial design, the optimization algorithm (GCMMA) can locate a better local optimum and achieve better instantaneous thermal performance.

Not only can the transient pseudo-3D model help to improve instantaneous performance but also reduce the pumping power consumption to some extent (66.7% of the reduction rate of pumping power in Section 5.3).

Steady-state optimization has been shown to be incapable of replacing the transient analysis if a better instantaneous thermal performance is desired.

The inlet conditions set in this study are constant with respect to time, which means that the model may not be well adapted to real transient conditions, such as oscillating inlet pressure and time-dependent thermal generation. Treating these conditions is the subject of future work and should be treated in a similar framework as long as the penalty factors are investigated for the truly transient conditions.

## Acknowledgements

The first and second author have been supported by Project of the Key Program of National Natural Science Foundation of China under the Grant Numbers 11572120.